# A Unified Unsupervised Framework for Genome-Wide Association Studies in Heterogeneous Populations

Xiong Shen[1,2†], Zhenshuang Tang[2†], Yong Liao[1], Zhiyuan Lin[1], Huajun Zhou[1], Haohao Zhang[3], Yangfan Liu[1], Dong Yin[1], Yue Wang[1], Yuan Quan[7], Zhuqing Zheng[8], Xiong Xiong[1,4], Yuhua Fu[1,5], Shuhong Zhao[1,2,4,5,6], Xinyun Li[1,4,5,6*], Lilin Yin[1,4,5,6*], Xiaolei Liu[1,4,5,6*]

[1] *Key Laboratory of Pig Digital and Intelligent Breeding, Ministry of Agriculture and Rural Affairs & College of Animal Science and Technology, Huazhong Agricultural University, Wuhan 430070, China;*

[2] *Yazhouwan National Laboratory, Sanya 572024, China;*

[3] *School of Computer Science and Technology, Wuhan University of Technology, Wuhan 430070, China;*

[4] *Hubei Hongshan Laboratory, Wuhan 430070, China;*

[5] *Key Laboratory of Agricultural Animal Genetics, Breeding and Reproduction, Ministry of Education, Wuhan 430070, China;*

[6] *Frontiers Science Center for Animal Breeding and Sustainable Production, Ministry of Education, Wuhan 430070, China;*

[7] *College of Informatics, Huazhong Agricultural University, Wuhan 430070, China;*

[8] *Institute of Food Nutrition and Health, Jingchu University of Technology, Jingmen, Hubei 448000, China;*

[†] *These authors contributed equally to this work;*

[*] *Corresponding authors: Xinyun Li (xyli@mail.hzau.edu.cn), Lilin Yin (ylilin@mail.hzau.edu.cn), Xiaolei Liu (xiaoleiliu@mail.hzau.edu.cn).*

## Abstract

Genome-wide association studies (GWAS) have greatly advanced the discovery of genetic variants underlying complex traits and diseases. Yet in heterogeneous populations, existing GWAS strategies typically either pool all individuals under an assumption of population homogeneity or perform meta-analysis across predefined subgroups, both of which are limited when latent genetic heterogeneity attenuates subgroup-specific effects and masks true associations or subgroup labels are imprecise. Here we present UCALM, a unified unsupervised framework that infers genetically homogeneous subgroups directly from the data and integrates subgroup-specific GWAS with a novel layered meta-analysis method to capture both shared and subgroup-specific association signals. Through extensive simulations and analyses of large-scale human and livestock cohorts, including the UK Biobank (n ≈ 487,000) and a heterogeneous pig cohort (n ≈ 85,000), we demonstrate that UCALM substantially alleviated the mean genomic inflation across 24 UK Biobank traits to 1.17 compared with 1.43 for GLM and 1.37 for LDAK-KVIK, and further revealed 74 loci in the pig cohort that were previously obscured by conventional approaches. Our results establish a robust and broadly applicable strategy for association mapping in structured populations and improve the resolution of genetic signals across diverse species.

## Introduction

Genome-wide association studies (GWAS) have played a significant role in revealing the genetic architecture of complex traits in humans, livestock and plants, identifying thousands of loci that contribute to phenotypic variation[1-5]. The expansion of biobanks and breeding programs has enabled the integration of datasets comprising hundreds of thousands of individuals from diverse genetic backgrounds[6-12]. Although multi-population cohorts can increase power to discover novel associations and refine causal variants, their utilization is complicated by population stratification and cryptic relatedness[13,14], particularly when genetic backgrounds are heterogeneous. Conventional GWAS strategies typically follow one of two paradigms: pooled analysis of all individuals under an assumption of population homogeneity, in which Principal Component Analysis (PCA) and Genomic Relationship Matrix (GRM) are used to correct for broad population structure and cryptic relatedness[15]; or meta-analysis across predefined subgroups, which relies heavily on prior subgroup labels that are sometimes unavailable, often subjectively defined and may not accurately reflect the underlying genetic structure[16]. However, the former may fail to capture fine-scale, non-linear genomic substructures, potentially leading to diluted subgroup-specific effects and masked association signals[17], whereas the latter may leave residual stratification that induces spurious associations.

Accurate statistical inference of population components directly from the data can provide an effective strategy for dissecting the hidden genetic architecture of complex populations, but the direct application of clustering algorithms to raw genotype data is fundamentally hindered by the curse of dimensionality inherent to sequencing data[18,19]. Existing approaches for population structure inference and genetic clustering include distance-based clustering methods, such as k-means, and model-based ancestry inference tools, such as STRUCTURE, ADMIXTURE and fastSTRUCTURE[20-22]. Although widely used, these methods typically require the number of clusters or ancestry components to be specified in advance and evaluated by comparing solutions across candidate values after repeated analyses. These limitations highlight the need for an unsupervised and adaptive clustering strategy that can infer fine-scale genetic subgroups with minimal manual intervention.

Determining the optimal association model for heterogeneous cohorts presents a significant challenge beyond population inference. Pooled analyses using Linear Mixed Models (LMMs) typically assume homogeneous genetic effects and estimate an average genetic effect across individuals, which can attenuate or mask heterogeneous signals[23,24]. By contrast, the meta-analysis strategies designed to integrate sub-population summary statistics can, in principle, accommodate population heterogeneity; however, existing methods each have distinct strengths and limitations, and no single approach performs well across all kinds of heterogeneous populations and genetic architectures. For example, Fixed-Effects (FE) models can suffer substantial power loss in the presence of heterogeneity[25-27], while Random-Effects (RE) models do not explicitly model ancestry-correlated effect patterns[28], and the Meta-Regression of Multi-Ancestry Genetic Association (MR-MEGA), although able to model linear ancestry gradients, may lack sensitivity to population-specific effects when association signals are restricted to a single subgroup, as regression-based modeling tends to dilute sharply localized effects across populations[29]. Together, these limitations underscore the need for a unified framework that can simultaneously capture shared, ancestry-correlated, and population-specific genetic effects.

To address these challenges, we developed UCALM (**U**nsupervised adaptive **C**lustering-based **A**ssociation and **L**ayered **M**eta-integration), a unified framework for association mapping in heterogeneous populations. UCALM first infers genetically homogeneous subgroups directly from genotype data in a fully data-driven and unsupervised manner, thereby eliminating the need for manual intervention or predefined subgroup labels that may be unavailable or fail to reflect the true underlying genetic structure. It then performs subgroup-specific GWAS to prevent heterogeneous effects from being diluted in pooled analyses and to improve the detection of subgroup-dependent association signals. Finally, UCALM integrates subgroup-level association evidence through a novel layered meta-analysis strategy, enabling the joint capture of shared, ancestry-correlated and subgroup-specific genetic effects that are difficult to resolve with any existing single meta-analysis model. By integrating data-driven subgroup inference, subgroup-aware association mapping and layered meta-integration, UCALM delivers greater statistical power and more robust Type I error control than existing mainstream methods across extensive simulations and real datasets, including the UK Biobank ($n \approx 487{,}000$) and a heterogeneous pig cohort ($n \approx 85{,}000$), providing a unified and powerful solution for resolving genetic heterogeneity and improving locus discovery ability in structured populations. The software and detailed user manual are freely available at https://mvp-cpp.github.io/tutorial/ucalm.

## Results

### Overview of the UCALM framework

The UCALM framework is composed of three major components: unsupervised subgroup inference from genotype data, subgroup-specific GWAS, and layered meta-integration of subgroup-level association signals (Fig. 1).

First, UCALM infers genetically homogeneous subgroups directly from genotype data through a fully automated unsupervised procedure that integrates principal component analysis (PCA), Uniform Manifold Approximation and Projection (UMAP)[17], and Hierarchical Density-Based Spatial Clustering of Applications with Noise (HDBSCAN)[30]. In this framework (Fig. 1a), PCs that are derived from GRM provide an initial low-dimensional representation of the high-dimensional genotype matrix, UMAP further resolves fine-scale non-linear relationships among individuals, and an optimized Auto-HDBSCAN delineates genetically homogeneous subgroups from the resulting manifold without requiring manual intervention or prior subgroup labels. This design enables UCALM to recover latent population structure while avoiding the subjectivity and potential misspecification associated with predefined subgroup assignment.

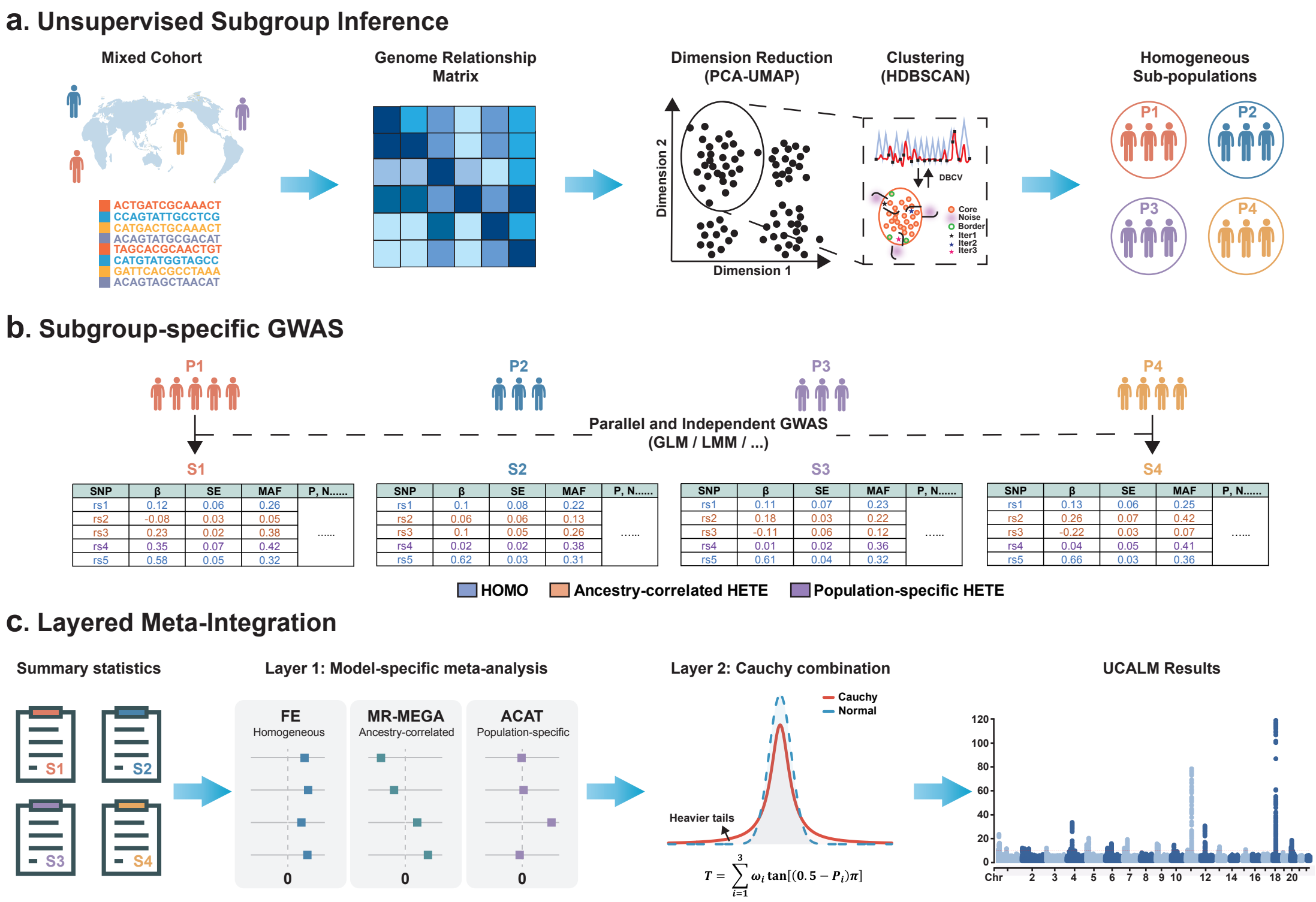

**Fig. 1 Overview of the UCALM framework. a,** Unsupervised subgroup inference. UCALM infers genetically homogeneous subgroups directly from genotype data using PCA, UMAP and Auto-HDBSCAN. **b**, Subgroup-specific GWAS. GWAS is performed independently within each inferred subgroup to obtain subgroup-level summary statistics (e.g., Effect size, SE, *P-value*). **c**, Layered meta-integration. FE, MR-MEGA and a subgroup-based ACAT component capture homogeneous, ancestry-correlated heterogeneous and population-specific effects, respectively; their *P* values combined in a second-layer Cauchy combination to generate the final UCALM association signal.

Second, UCALM performs GWAS within each inferred subgroup, thereby preventing heterogeneous genetic effects from being diluted in pooled analyses and enhancing the detection of subgroup-dependent association signals (Fig. 1b). This divide-and-conquer architecture is compatible with a broad spectrum of statistical models, including standard GLM and LMM, as well as computationally efficient algorithms such as BOLT-LMM, fastGWA, LDAK-KVIK, and REGENIE[14,15,23,24,31-33].

Lastly, UCALM integrates subgroup-level summary statistics through a layered meta-analysis framework (Fig. 1c). In the first layer, UCALM applies fixed-effects meta-analysis (FE)[25], Meta-Regression of Multi-Ancestry Genetic Association (MR-MEGA)[29], and a population-specific test based on the Aggregated Cauchy Association Test (ACAT)[34,35]. These three complementary association components are designed to capture shared effects, ancestry-correlated heterogeneous effects, and subgroup-specific effects, respectively. Their resulting statistics are then combined in a second layer using an ACAT-based omnibus combination to generate the final association signal. By hierarchically aggregating complementary evidence across models, UCALM derives a unified association measure that is robust to statistical dependence and effectively captures a broad spectrum of allelic effect architectures.

**Fine-scale characterization of genetic structure across species**

In our proposed UCALM framework, principal component analysis can effectively reduce the dimensionality of high-density SNP data and mitigate the curse of dimensionality; however, determining the optimal number of principal components (PCs) to retain is a critical issue, as it directly affects downstream UMAP embedding and the resolution of population subgrouping. We therefore systematically evaluated the effects of varying numbers of PCs on clustering accuracy and stability in two independent cohorts comprising human and pig populations, using the concordance between unsupervised adaptive clustering results and available sample prior labels. These analyses showed that retaining the first 10 PCs enables robust characterization of both global population genetic architecture and fine-scale subpopulation differentiation, while yielding clustering assignments with acceptable concordance to prior sample labels (Supplementary Figs. 1-2). Notably, prior sample labels often fail to capture the full complexity of underlying population structure and should therefore be regarded only as a secondary reference for selecting the optimal number of PCs. Furthermore, retaining too few PCs obscures subtle subpopulation structure, whereas including too many leads to degradation of distance metrics in high-dimensional space due to the curse of dimensionality. Accordingly, all downstream analyses in this study were based on clustering results generated from the first 10 PCs.

We assessed the scalability and versatility of the adaptive clustering framework across two large-scale real datasets characterized by divergent demographic histories comprising the continuous admixture of human populations and the discrete selection-driven stratification of livestock, both of which included predefined sample labels that enabled assessment of concordance between unsupervised clustering assignments and prior population classifications. Within the UK Biobank (UKB; n = 487,160), our proposed framework resolved global ancestral gradients alongside fine-scale substructures in a fully unsupervised manner (Fig. 2a, b). The identification of 20 distinct clusters, with full cluster assignment details against self-reported ancestry provided in Supplementary Data 1, exhibited substantial concordance between inferred subgroup assignments and manually curated population labels, particularly at the level of major regional ancestries across different geographic groups. Notably, however, UCALM further resolved multiple fine-scale substructures within several broadly defined populations, most prominently within the British cohort, indicating that substantial cryptic genetic heterogeneity remains masked under conventional label definitions. In addition, individuals annotated as “Unknown” in the original metadata were robustly assigned into genetically coherent subgroups by UCALM, demonstrating its ability to recover latent population structure even in the absence of reliable prior information.

We further evaluated the performance of UCALM in a heterogeneous pig cohort comprising multiple breeds and commercial lines derived from different breeding companies (n = 85,594; Figs. 2c, d). Overall, the inferred subgroup assignments showed good concordance with manually defined breed and line annotations, demonstrating the ability of UCALM to accurately resolve major genetic structure (i.e., Duroc, Landrace, Yorkshire, and Pietrain) within complex livestock populations. The full clustering results for breed and within-breed line assignments are detailed in Supplementary Data 2. Importantly, UCALM additionally identified previously unrecorded sub-line structures within several broadly defined populations, revealing cryptic genetic stratification that was not captured by existing metadata. Moreover, individuals originating from different companies but sharing similar genetic backgrounds were consistently grouped into the same genetically coherent clusters, indicating that UCALM captures underlying genomic relationships rather than artificial organizational boundaries.

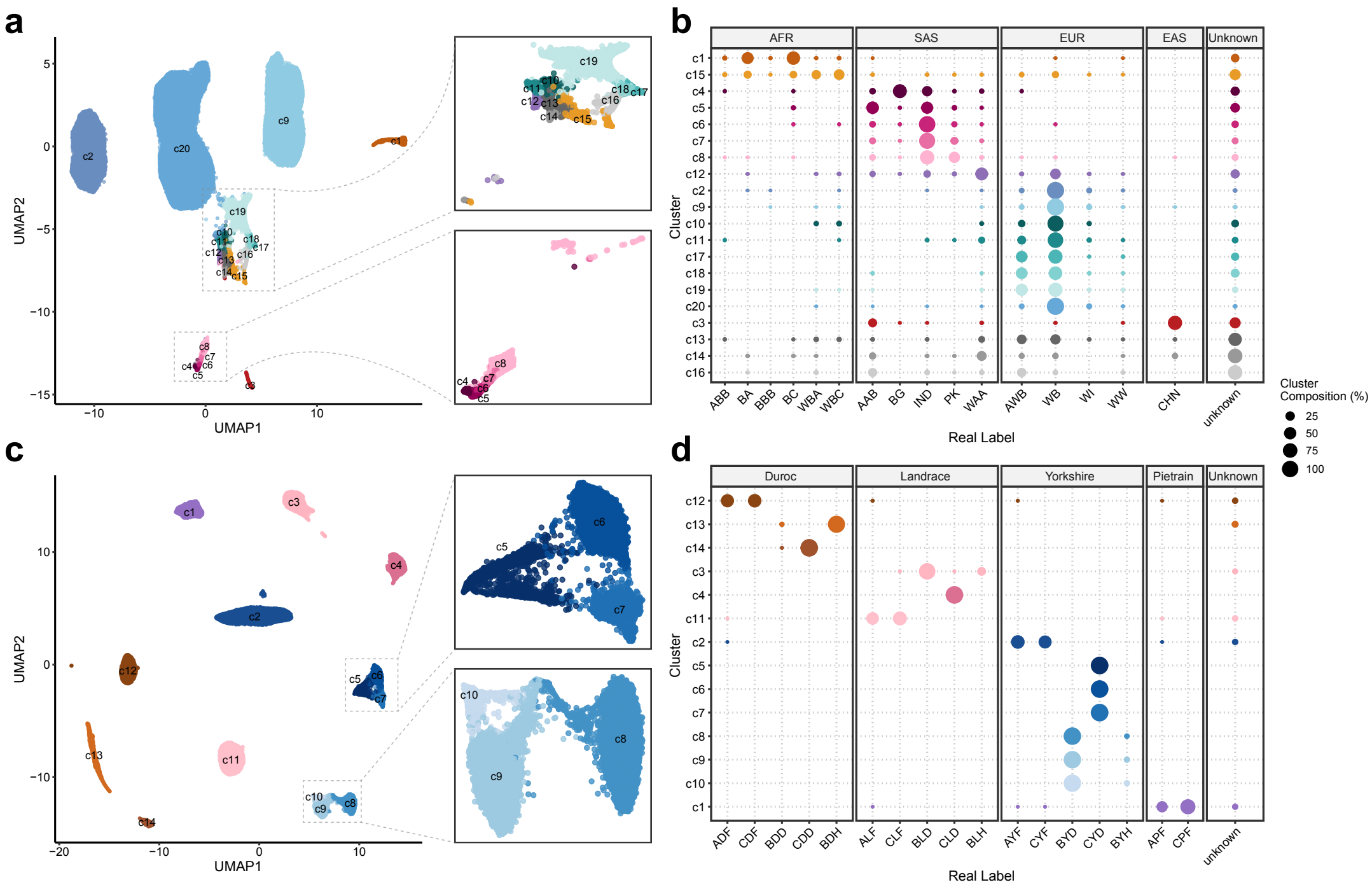

**Fig. 2 Inference of population structure in human and pig cohorts.** **a**, UMAP projection of the UK Biobank cohort (n = 487,160), colored by 20 clusters inferred by UCALM. **b**, Concordance between the inferred clusters and self-reported ancestry labels. **c**, UMAP projection of the pig cohort (n = 85,594), colored by 14 clusters inferred by UCALM. **d**, Correspondence between the inferred pig clusters and reference population labels. In **b** and **d**, dot size represents the proportion of individuals within each cluster assigned to the corresponding reference label, and dot colors match the cluster colors in **a** and **c**. Column facets group labels by major ancestry in **b** and breed in **d**. SAS, South Asian; EUR, European; AFR, African; EAS, East Asian. In **b**, the x-axis shows UK Biobank

self-reported ancestry subcategories, with full code definitions are provided in Supplementary Data 1. In **d**, the x-axis shows pig population codes, the first letter denotes company (A, B or C), the second letter denotes breed (D, Duroc; L, Landrace; Y, Yorkshire; P, Pietrain), and the third letter denotes line (F, French; D, Danish; H, Chinese).

As illustrated in either Fig. 2b for human or Fig. 2d for pig populations, horizontal comparisons show that UCALM-derived subgroups often encompass multiple manually predefined labels, indicating that conventional label-based stratification may artificially fragment populations with highly similar genetic backgrounds into smaller groups. Such fragmentation reduces effective sample size within subpopulations and consequently leads to a loss of statistical power in downstream association analyses. In contrast, vertical comparisons reveal that manually defined labels are frequently too coarse to capture fine-scale genetic structure, with multiple genetically distinct subgroups often coexisting within a single predefined population. This hidden substructure may introduce residual confounding and increase the risk of false-positive associations in conventional GWAS frameworks. Collectively, these findings demonstrate that UCALM, by inferring population structure directly from genomic data in a fully data-driven manner, provides substantially higher resolution and biological fidelity than manually curated labels. These results further highlight the accuracy, robustness and flexibility of UCALM for fine-scale population stratification across diverse populations and species.

To evaluate the computational efficiency and scalability of the UCALM in subgroup inference, we performed a benchmark analysis using six randomly sampled subsets from the human cohort of approximately 480,000 individuals, with sample sizes of 15,000, 30,000, 60,000, 120,000, 240,000 and 480,000. The workflow included three core steps: computation of the first 10 principal components, UMAP dimensionality reduction and Auto-HDBSCAN clustering. We also recorded the total running time of the full program, including data loading, preprocessing and output generation. As shown in Table 1, for the full-scale dataset of 480,000 individuals, the total running time of the full program was 1,714 seconds (28.6 minutes), with a peak memory usage of 929.71 GB. PCA, UMAP and Auto-HDBSCAN required 81, 353 and 315 seconds, respectively. Among the three core modules, UMAP was one of the main time-consuming steps, increasing from 24 seconds at 15,000 samples to 353 seconds at 480,000 samples. Auto-HDBSCAN also showed a marked increase in running time at large sample sizes, from 2 seconds at 15,000 samples to 315 seconds at 480,000 samples. The PCA step only computed the first 10 principal components and used truncated eigenvalue decomposition to avoid the high computational cost of full matrix decomposition, requiring only 81 seconds for 480,000 samples. As the sample size increased from 15,000 to 480,000, the total running time of the full program increased from 30 to 1,714 seconds, demonstrating that the workflow remains computationally tractable at biobank scale on a high-performance server. Overall, these results highlight the super computational efficiency and support the engineering feasibility of the UCALM in subgroup inference for large-scale biobank and breeding population datasets.

**Table 1. The computational performance including running time (in seconds) and memory consumption (GB) of the automated unsupervised population clustering procedure.**

| Population size | PCA | UMAP | Auto-HDBSCAN | Total (s) | Memory (GB) |
|---|---|---|---|---|---|
| **15,000** | 1 | 24 | 2 | 30 | 1.25 |
| **30,000** | 1 | 32 | 4 | 42 | 5.02 |
| **60,000** | 2 | 52 | 10 | 85 | 20.11 |
| **120,000** | 6 | 74 | 26 | 197 | 80.47 |
| **240,000** | 22 | 215 | 78 | 644 | 321.87 |
| **480,000** | 81 | 353 | 315 | 1714 | 929.71 |

Note: All benchmarks were performed on a high-performance server equipped with two 2.70 GHz Intel Xeon 8485P processors, using 64 CPU cores for computation. The PCA step computed only the first 10 principal components. Running time is reported in seconds, and memory usage is reported in GB as peak memory usage. PCA, UMAP and Auto-HDBSCAN indicate the running time of the three core modules. Total indicates the running time of the full program, including data loading, preprocessing, core module execution and output generation, and therefore is not the arithmetic sum of the three core module times.

## Statistical power and type I error control

To further evaluate the statistical performance of UCALM in GWAS, we constructed a genetically heterogeneous simulation panel with representation across major ancestries by sampling 50,000 individuals from the UK Biobank, including 35,000 European (EUR), 7,000 African (AFR), 6,600 South Asian (SAS) and 1,400 East Asian (EAS) individuals, and simulated phenotypes under four genetic effect architectures with increasing levels of complexity. Three heritability levels (0.2, 0.5 and 0.8) were considered. According to the cross-population distribution of marker effects, simulation scenarios were classified into four categories: HOMO, in which causal effects were homogeneous across populations; HOMO_HETE, in which homogeneous and heterogeneous effects coexisted; HETE, in which marker effects were entirely ancestry-correlated heterogeneous across populations; and HETE_onlyEAS, in which causal effects were present exclusively in the East Asian (EAS) subgroup and absent in all other populations. Detailed simulation settings, including the number and effect sizes of causal variants and their genomic distributions, are described in the Methods section. UCALM was compared against conventional linear mixed models (LMMs) and several meta-analysis approaches, including fixed-effects (FE), random-effects (RE) and MR-MEGA models, for which subgroup labels required for meta-analysis were inferred using UCALM. Type I error rates were assessed using the average equivalent $\chi_1^2$ statistic of null SNPs and the false positive rate (FPR) at significance thresholds of $P < 0.05$, $5 \times 10^{-3}$, $5 \times 10^{-4}$, and $5 \times 10^{-5}$.

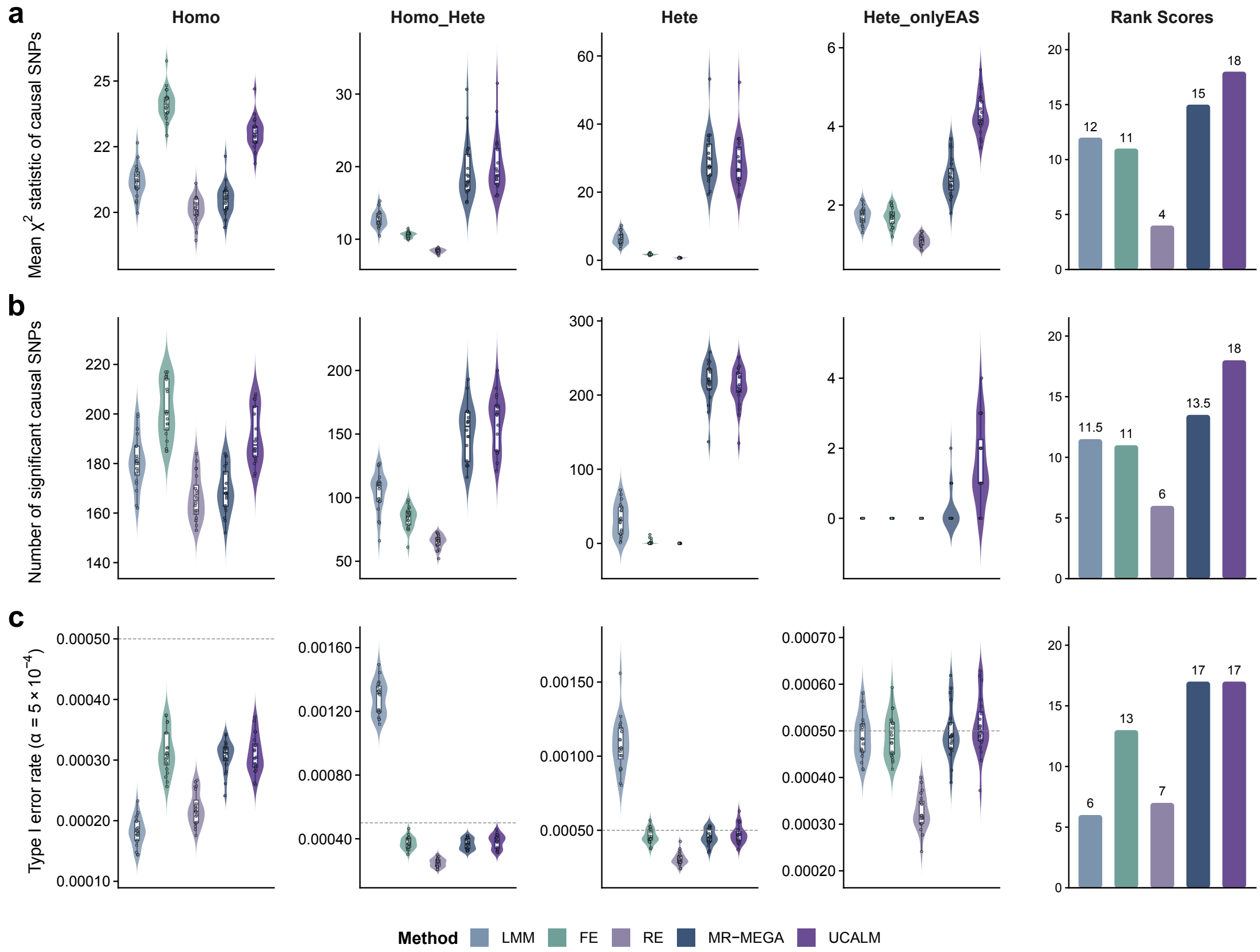

**Fig. 3 Benchmarking statistical power and type I error control across four genetic architectures.** Performance comparison of five methods (LMM, FE, RE, MR-MEGA, and UCALM) based on 20 independent simulation replicates with heritability $h^2 = 0.5$. Columns correspond, from left to right, to homogeneous effects (HOMO), mixed homogeneous and heterogeneous effects (HOMO_HETE), ancestry-correlated heterogeneous effects (HETE), and population-specific effects confined to the EAS subgroup (HETE_onlyEAS). **a,** Distribution of the mean $\chi^2$ statistic for all causal variants, quantifying the overall strength of association signals. **b,** Empirical statistical power, measured as the number of causal variants reaching the Bonferroni significance threshold. **c,** Empirical type I error rates assessed using null variants at a nominal significance threshold of $P < 5 \times 10^{-4}$. The grey dashed line in **c** indicates the expected nominal error rate. The 'Rank scores' is the sum of ranked position of five methods in the four simulated scenarios, higher scores indicate better performance of the method, with a maximum score of 20 points. When methods showed identical performance, they were assigned the average of the corresponding ranked scores.

As shown in Fig. 3, we can conclude that UCALM demonstrated the best statistical power (Supplementary Tables 5-6) while maintaining effective type I error control across all simulated architectures and heritability levels (Supplementary Figs. 5-7). The mean $\chi_1^2$ for null variants were generally well calibrated across all five methods: FE and MR-MEGA yielded values close to 1.00, whereas UCALM, LMM, and RE showed slightly conservative calibration under some scenarios (Supplementary Table 1).

Quantitative evaluation of the pooled false-positive rates further supported these trends (Supplementary Tables 2-4). The conservative calibration of UCALM likely reflects its robust control of residual population stratification and confounding, particularly in genetically heterogeneous populations.

Under the homogeneous-effect scenario (HOMO), UCALM performed comparably to the most powerful fixed-effects (FE) model, while substantially outperforming LMM and RE (Fig. 4d). Under the ancestry-correlated heterogeneous scenario (HETE), UCALM and MR-MEGA achieved the highest power (Fig. 3b), whereas pooled LMM frequently produced inflated false-positive signals, reflecting the inability of homogeneous-effect models to adequately account for cross-population effect heterogeneity (Fig. 4b). The advantage of UCALM was most pronounced under the population-specific architecture (HETE_onlyEAS), where causal effects were confined to the East Asian subgroup. In this specific architecture, UCALM substantially outperformed alternative approaches, and notably, LMM, FE and RE failed to detect any significant loci (Fig. 4a), whereas MR-MEGA showed only limited sensitivity to effects restricted to minority sub-populations. Within the mixed architecture (HOMO_HETE), UCALM provided the most balanced performance by capturing both shared and stratified loci (Fig. 4c). All the above trends remained consistent across lower ($h^2 = 0.2$) and higher ($h^2 = 0.8$) heritabilities (Supplementary Figs. 3-4), which demonstrates the stability of UCALM across varying genetic effect magnitudes.

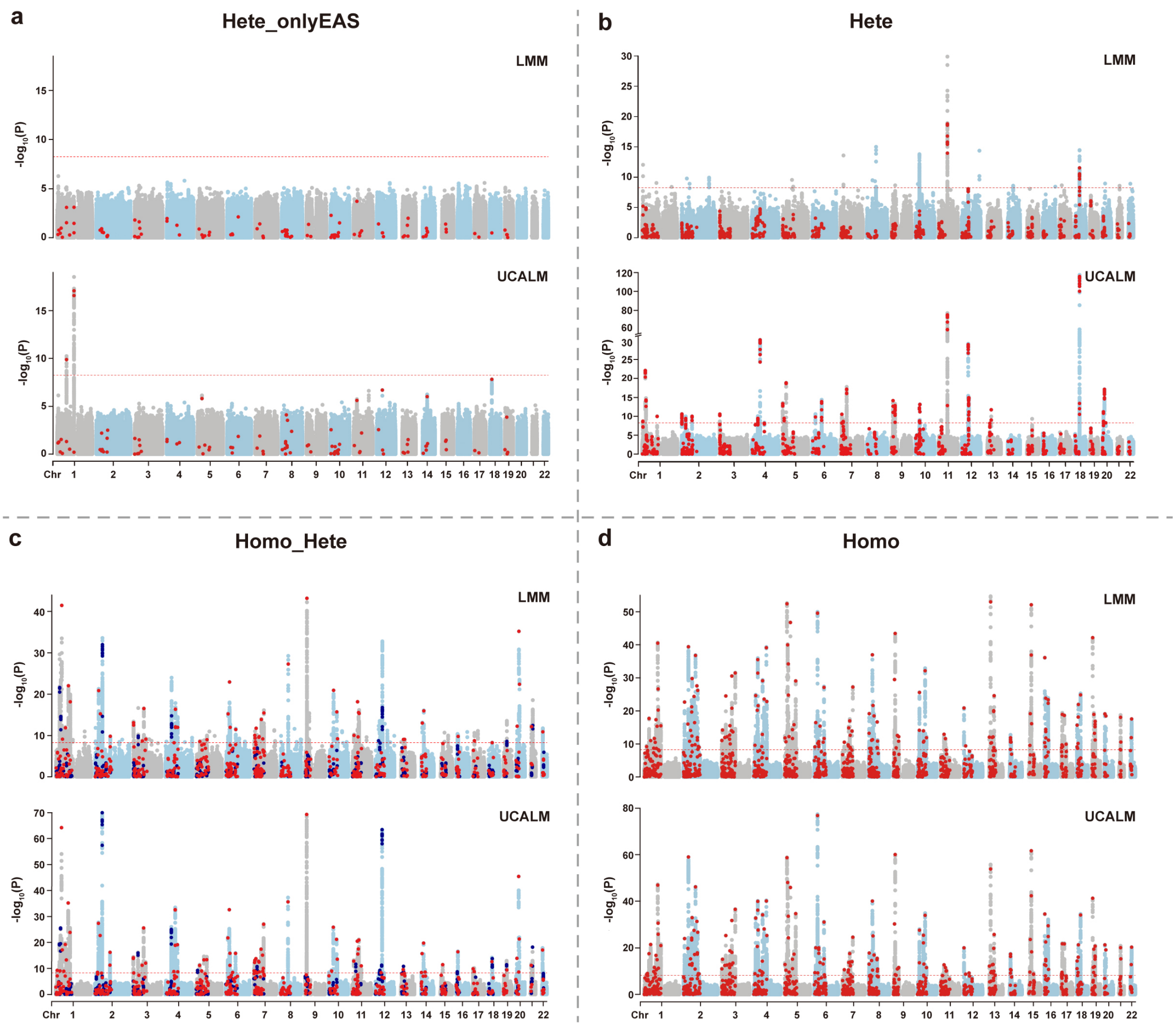

**Fig. 4 Comparison of association signals across multiple architectures.** Representative Manhattan plots displaying association statistics ($-\log_{10}P$) from the standard Linear Mixed Model (LMM, top) and UCALM (bottom) under varying genetic architectures ($h^2 = 0.5$). **a,** Population-specific heterogeneity (HETE_onlyEAS), in which causal effects are confined to the East Asian population. **b,** Ancestry-correlated heterogeneity (HETE), in which causal effects vary linearly with population differentiation. **c,** Mixed architecture (HOMO_HETE), containing causal SNPs with homogeneous and heterogeneous effects **d,** Homogeneous effects (HOMO). In **a**, **b**, and **d**, red dots indicate simulated causal SNPs. In **c**, red and dark blue dots indicate causal SNPs with homogeneous and heterogeneous effects, respectively. The horizontal dashed line denotes the Bonferroni significance threshold ($P < 5.56 \times 10^{-9}$).

## Application of UCALM to 24 quantitative traits in the UKB

### *Rigorous control of genomic inflation with comparable statistical power*

To evaluate the performance of UCALM across diverse genetic architecture of real dataset, we first applied UCALM to 24 quantitative traits in the UK Biobank (UKB) cohort (n ≈ 480,000). Although this cohort is predominantly of European ancestry (approximately 90%), it still contains fine-scale substructure and cryptic relatedness, thereby providing a rigorous test bed for evaluating statistical models in near-homogeneous populations. We compared the performance of UCALM with both a conventional general linear model (GLM) incorporating PCA covariates and the LDAK-KVIK, a computationally efficient approach that is statistically equivalent to a linear mixed model (LMM).

Across the 24 traits analyzed in the UK Biobank, association signals detected by UCALM were highly concordant with those identified by conventional GLM and LDAK-KVIK, despite the markedly finer subgroup structure inferred by UCALM, as shown in the clustering results of Fig. 2a, b. Representative Manhattan plots for standing height (HT; n = 485,495) and body mass index (BMI; n = 484,961) further illustrated this overall consistency across methods (Fig. 5a, b). These findings suggest that the major association signals in UKB are robust and largely preserved across analytical frameworks. A plausible explanation is that, although UCALM resolves subtle subgroup structure, the UKB cohort remains relatively homogeneous at a broad scale owing to its predominantly European ancestry, such that major association signals are consistently captured by all methods. In this setting, the principal advantage of UCALM lies not in altering the overall landscape of association results, but in improving control of residual structure and calibration without compromising signal recovery.

We rigorously assessed control of population stratification by quantifying the genomic inflation factor ($\lambda_{GC}$) across distinct minor allele frequency categories for each trait (Supplementary Table 7). All evaluated methods showed global $\lambda_{GC}$ values greater than 1.0. Such inflation is expected in large-scale analyses of highly polygenic traits and is primarily attributable to polygenicity and substantial sample sizes rather than inadequate correction for population stratification[23,36,37]. Nevertheless, UCALM consistently demonstrated superior frequency calibration and yielded the lowest global inflation across the 24 traits (Fig. 5c). The mean global $\lambda_{GC}$ for UCALM was 1.1745, which is substantially lower than that observed for GLM at 1.4325 and LDAK-KVIK at 1.3716. For traits with high heritability such as standing height, UCALM markedly attenuated the global inflation $\lambda_{GC}$ from 1.8096 in the GLM and 1.8569 in the LDAK-KVIK to 1.5837. Stratified analyses further showed that UCALM provided the most conservative rare-variant calibration, with a mean rare-variant $\lambda_{GC}$ of 0.9159, compared with 1.1780 for GLM and 1.0750 for LDAK-KVIK, while preserving the expected polygenic inflation among common variants (Supplementary Table 7). Together with the simulation evidence of effective type I error control in heterogeneous settings (Fig. 3c), the reduced $\lambda_{GC}$ values observed in the UKB application suggest that UCALM improves control of residual population structure and cryptic relatedness without compromising major association signals.

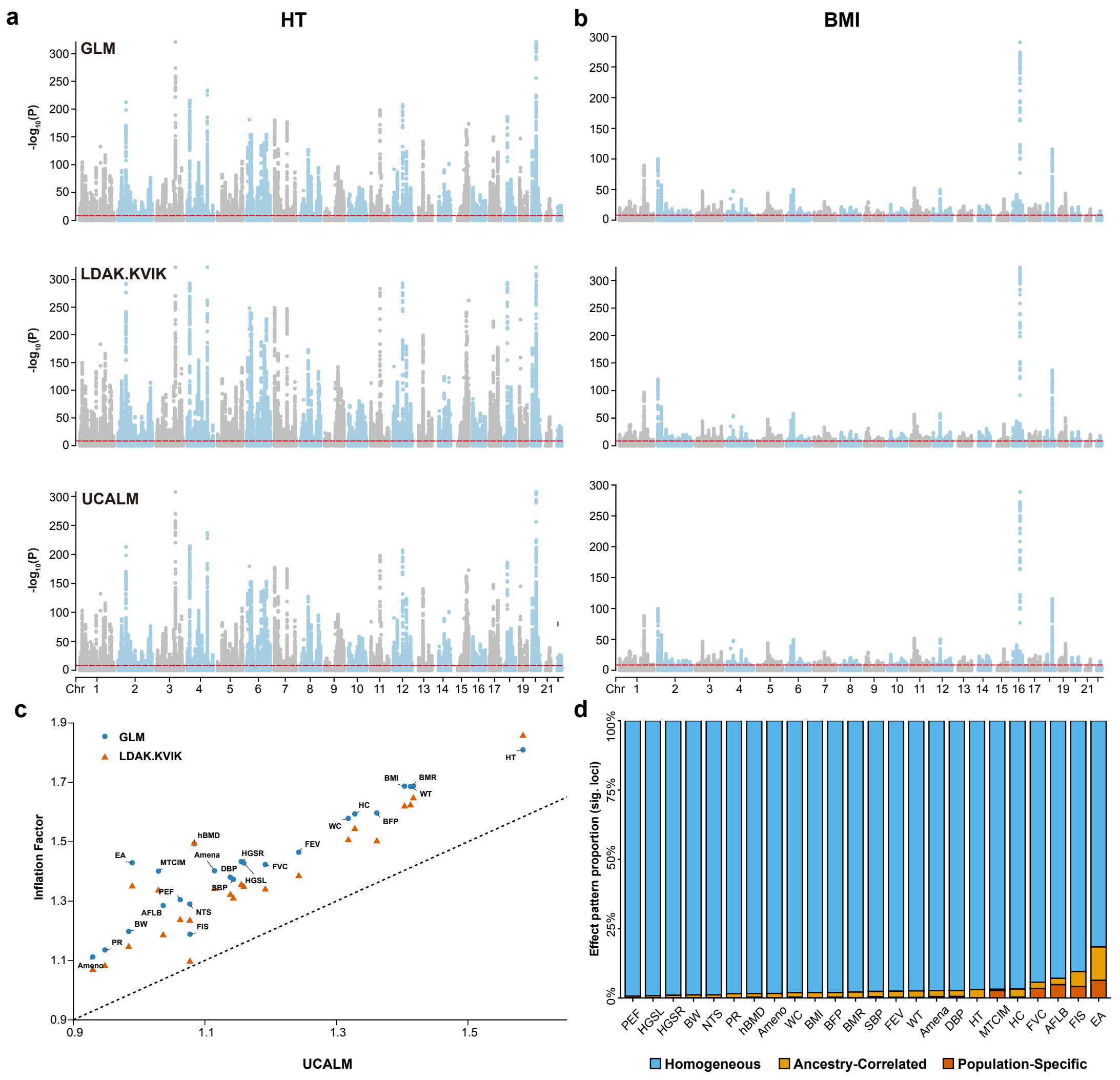

**Fig. 5 Association results, genomic inflation and effect architectures for 24 UK Biobank quantitative traits. a, b,** Manhattan plots for standing height (HT; *n* = 485,495) (**a**) and body mass index (BMI; *n* = 484,961) (**b**) generated using GLM, LDAK-KVIK and UCALM. Rows indicate methods; the *x* axis represents chromosomal position, and the *y* axis indicates statistical significance on a $-log_{10}(P)$ scale. The red dashed line denotes the genome-wide significance threshold ($P < 3.51 \times 10^{-9}$; 14,265,293 tested variants). **c**, Comparison of genomic inflation factors between UCALM and GLM or LDAK-KVIK across 24 traits. The dashed diagonal line denotes equal inflation. **d**, Proportions of significant loci assigned to homogeneous, ancestry-correlated and population-specific effect patterns. Trait abbreviations are defined in Supplementary Table 9.

## *Distinct profiles of homogeneous and heterogeneous effects across human complex traits*

We systematically dissected the genetic architecture underlying 24 quantitative traits by analyzing a component contribution of the significant loci identified by UCALM. Using a minimum *P*-value based classification strategy, each locus was assigned to its dominant effect category according to the most significant first-layer component statistic from the layered meta-analysis framework. Across traits, homogeneous effects emerged as the predominant driver, accounting for the vast majority (>95%) of associations in canonical anthropometric and physiological traits, including peak expiratory flow (PEF) and hand grip strength (HGSL/HGSR) (Fig. 5d).

Distinct patterns of allelic effects were observed in complex behavioral and life-history traits. In particular, traits such as fluid intelligence score (FIS) and educational attainment (EA) showed enrichment of ancestry-correlated and population-specific heterogeneity components (Fig. 5d). This pattern suggests that the genetic basis of these cognitive and socioeconomic traits may be more susceptible to ancestry-correlated or population-specific variation in effect size, potentially reflecting gene-by-environment interactions ($G \times E$) or divergent phenotypic definitions across sub-populations. These findings highlight the importance of the UCALM omnibus framework for capturing heterogeneous genetic effects that may be attenuated or missed under conventional homogeneous models.

**Application of UCALM to four economic traits in pigs**

***UCALM controls population stratification with significantly improved statistical power***

We next evaluated UCALM across multiple traits in a pig cohort comprising multiple breeds from different commercial sources, providing a stringent test bed for association mapping in a genetically heterogeneous population shaped by complex breeding histories, artificial selection and pronounced population structure. Analysis of the genomic inflation factor stratified by minor allele frequency revealed clear differences among GLM, LMM and UCALM (Fig. 6c; Supplementary Table 8). GLM showed severe inflation across all four traits, with a mean overall $\lambda_{GC}$ of 13.7018, whereas both LMM and UCALM effectively controlled genomic inflation. For traits such as backfat thickness adjusted to 100 kg (BF) and loin muscle area adjusted to 100 kg (LMA), LMM showed deflation among common variants, consistent with potential overcorrection when principal components and the genomic relationship matrix are jointly included. By contrast, UCALM mitigated this loss of power by conducting subgroup-specific GWAS within relatively homogeneous genetic backgrounds. UCALM also provided substantially improved control of genomic inflation for rare variants than LMM. In genetically complex populations, rare variants are more likely to be population-specific, and LMM appeared unable to fully account for this structure, resulting in inflation for traits such as FCR and LMA.

Having established that UCALM provides improved control of false-positive signals, we next evaluated whether this improved calibration was accompanied by gains in discovery power. We systematically quantified the concordance among methods and the additional discoveries enabled by UCALM using a genome-wide window-based intersection analysis across all four economically important traits. The genome was partitioned into independent 100-kb windows as described in Methods, and then detection sensitivity was compared across five statistical approaches: Linear Mixed Model (LMM), Fixed-Effects (FE), Random Effects (RE), MR-MEGA, and UCALM. As shown in Fig. 6d and Supplementary Figs. 8-11, UCALM exhibited strong discovery power, capturing 126 loci shared across all five evaluated methods, and thereby confirming its reliability in recovering consensus association signals. Beyond these common signals, UCALM remained highly sensitive to loci identified by rigorous meta-analytic strategies, identifying 111 loci shared with FE and MR-MEGA but missed by both LMM and RE. Moreover, 61 loci were detected by both UCALM and MR-MEGA but not by either FE or LMM. These loci likely harbor ancestry-correlated effects, in which allelic effect sizes vary with breed differentiation, and would therefore be attenuated or overlooked in pooled analyses. Most importantly, UCALM uniquely identified 74 loci that were not detected by any other method. This additional discovery power is consistent with the contribution of the ACAT component, which is particularly sensitive to subgroup-specific effects, together with the integration of complementary association components. These results highlight that the strength of UCALM lies in its explicit modelling and integration of multiple classes of genetic association signals, thereby enabling the recovery of loci missed by conventional single-model approaches.

Similarly, we classified each significant locus according to its predominant effect architecture, distinguishing homogeneous, ancestry-correlated heterogeneous and subgroup-specific heterogeneous signals, consistent with the classification framework used in the preceding human analysis. As shown in Fig. 6e, homogeneous effects accounted for the majority of associations for traits such as AGE, whereas the FCR was dominated by signals exhibiting population-specific heterogeneity. Such marked architectural variation highlights the severe limitations of relying on a single statistical model in genetically complex pig populations shaped by intensive selection and structured breeding. Collectively, these results indicate that UCALM enhances candidate locus discovery by achieving an effective balance between statistical power and control of genetic heterogeneity.

***UCALM preserves shared signals while resolving subgroup-dependent loci for pig economic traits***

We next compared association results obtained using GLM, LMM and UCALM to determine whether UCALM improved the recovery of association signals in the heterogeneous pig cohort while retaining major loci detectable by conventional analysis (Fig. 6a, b; Supplementary Fig. 12). Across backfat thickness adjusted to 100 kg (BF), feed conversion ratio (FCR), age adjusted to 100 kg (AGE) and loin muscle area adjusted to 100 kg (LMA), UCALM recovered several established or broadly shared signals with clear biological relevance, including *MC4R* for BF ($P = 1.15 \times 10^{-63}$), a regulator of appetite and energy homeostasis[38], *VRTN* for BF ($P = 2.76 \times 10^{-87}$), a locus involved in vertebral development and body conformation[39], *INS/IGF2* region for FCR ($P = 2.94 \times 10^{-28}$), a locus previously implicated in pig growth and muscle development[40], *MC4R* for AGE ($P = 7.22 \times 10^{-59}$), and *VRTN* for LMA ($P = 5.60 \times 10^{-64}$) (Supplementary Data 3-6). Because GLM exhibited severe genomic inflation, candidate loci were classified relative to the standard LMM results. Candidate loci were annotated from lead SNPs in significant 100-kb windows and classified as common or novel according to whether the corresponding window also contained a genome-wide significant LMM signal. Importantly, UCALM also identified multiple candidate loci classified as novel in the annotation results, including *MRAP2* ($P = 4.17 \times 10^{-12}$), *MYOT* ($P = 2.11 \times 10^{-15}$) and *NEK1* ($P = 7.75 \times 10^{-32}$; Supplementary Fig. 13) for BF, *BRSK2* for FCR ($P = 1.94 \times 10^{-12}$), *ADAMTS6* for AGE ($P = 2.11 \times 10^{-9}$), and *TMEFF2* ($P = 1.16 \times 10^{-9}$) and *MYOD1* ($P = 4.78 \times 10^{-10}$) for LMA. Several of these loci have functions consistent with the corresponding traits, including *MRAP2* in *MC4R* signaling[41], *BRSK2* in glucose-stimulated insulin secretion[42], *ADAMTS6* in musculoskeletal biology[43], and *MYOD1* in myogenic differentiation[44]. *NEK1*, a kinase involved in primary cilium biology[45], represents a candidate locus for BF that warrants further functional investigation. These results indicate that UCALM preserves robust shared association signals while increasing sensitivity to loci that may be attenuated in pooled LMM analysis.

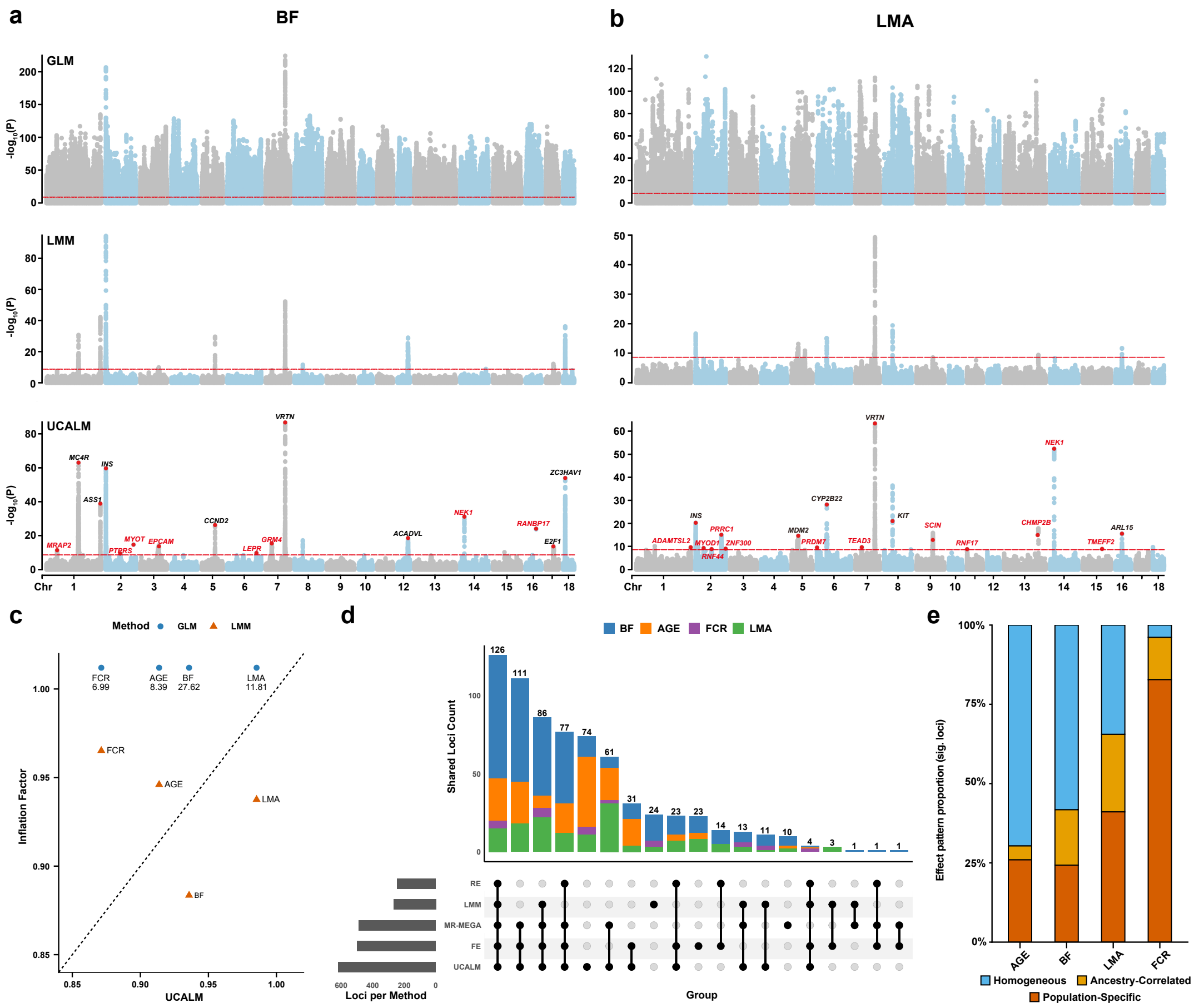

**Fig. 6 Comparison of genome-wide association results, genomic inflation, and effect patterns across four pig economic traits. a, b,** Manhattan plots comparing association results obtained using GLM, LMM, and UCALM for Backfat Thickness adjusted to 100 kg (BF, **a**) and Loin Muscle Area adjusted to 100 kg (LMA, **b**). The x-axis indicates chromosomal position, and the y-axis shows association significance as $-\log_{10}(P)$. The red horizontal dashed line indicates the genome-wide significance threshold ($P < 2.54 \times 10^{-9}$). Candidate genes are labelled in the UCALM panels; black labels denote loci detected by both LMM and UCALM, whereas red labels denote loci detected only by UCALM. **c,** Comparison of genomic inflation factors ($\lambda_{GC}$) between UCALM and GLM or LMM across BF, age adjusted to 100 kg (AGE), feed conversion ratio (FCR), and LMA. **d,** UpSet plot showing the overlap of significant 100-kb genomic windows detected by five association strategies (RE, LMM, MR-MEGA, FE, and UCALM) across the four traits. Vertical bars indicate intersection sizes, colors denote traits, and horizontal bars show the total number of windows detected by each method. **e,** Stacked bar plot showing the proportion of significant loci assigned to homogeneous, ancestry-correlated, and population-specific effect patterns for each trait. The y-axis indicates the percentage of significant loci within each trait, and colors denote the three effect-pattern categories.

Representative regional analyses further illustrated how UCALM resolves different effect patterns in the structured pig cohort (Fig. 7). The *MC4R* locus for BF represented a shared effect pattern: both LMM and UCALM detected a strong association signal, showing that UCALM retains major loci that are broadly detectable across the cohort. By contrast, the *TMEFF2* locus for LMA represented an ancestry-correlated effect pattern. This signal was not detected by LMM but became evident under UCALM, with subgroup-level estimates showing effects that varied across genetic backgrounds. The *MYOT* locus for BF represented a population-specific effect pattern, in which the association was mainly driven by specific genetic subgroups and was weak or absent in others. The *NEK1* locus provided an additional example for BF, reaching genome-wide significance with UCALM but not with LMM (Supplementary Fig. 13). Subgroup-level estimates revealed a combination of ancestry-correlated heterogeneous and population-specific effect patterns. Together, these examples show that UCALM preserves shared signals while recovering ancestry-correlated and population-specific loci that can be weakened or missed by pooled LMM analysis.

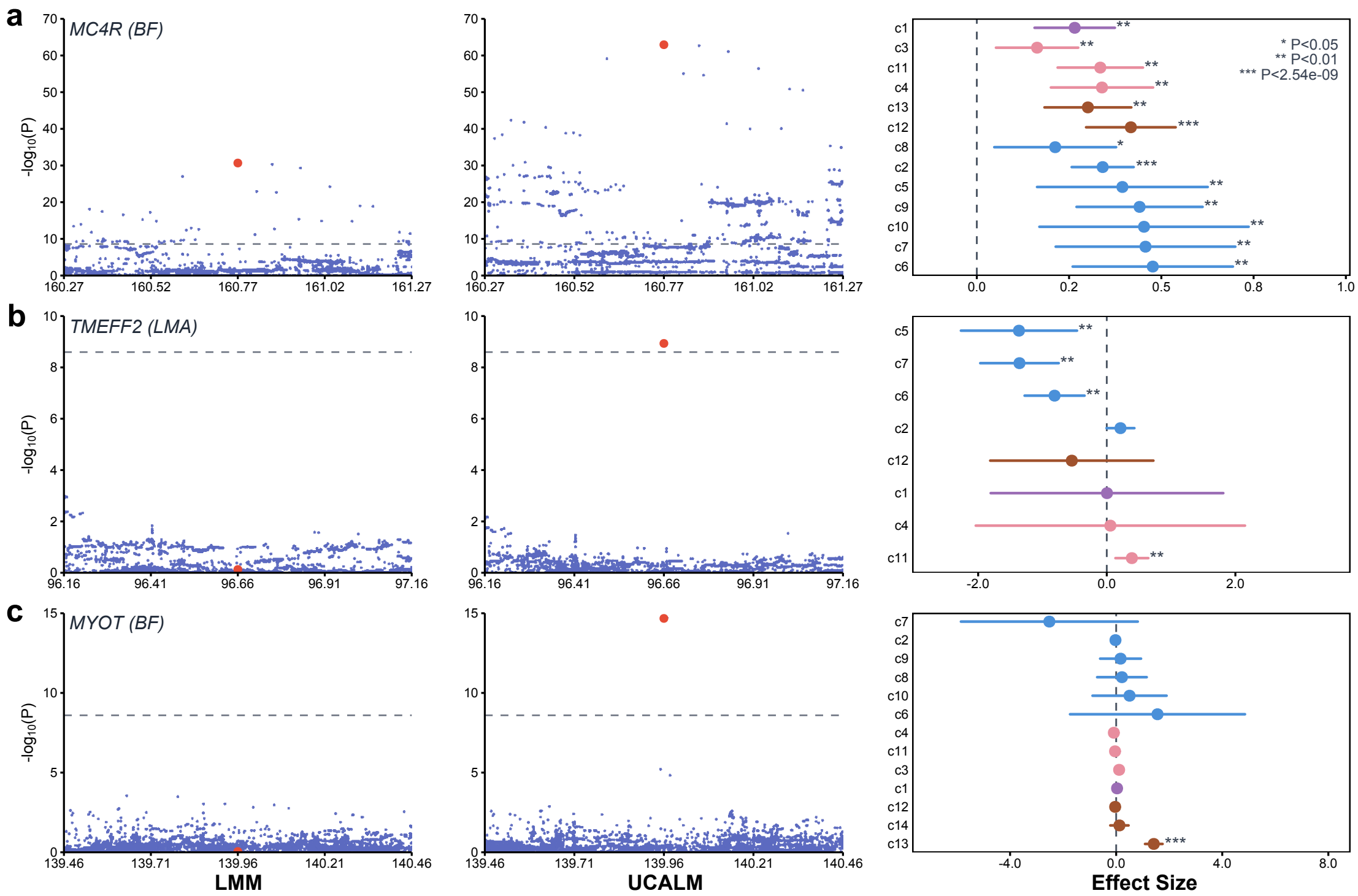

**Fig. 7 Representative regional associations and subgroup effect estimates for selected pig trait loci. a-c**, Regional association plots comparing LMM and UCALM signals at *MC4R* for BF (**a**), *TMEFF2* for LMA (**b**) and *MYOT* for BF (**c**). The red point denotes the lead SNP, and the horizontal dashed line indicates the genome-wide significance threshold ($P < 2.54 \times 10^{-9}$). The right panels show subgroup-level effect estimates for the lead SNP, with points indicating estimated effects and horizontal lines indicating 95% confidence intervals. Point and confidence-interval colors denote the predominant breed composition of each UCALM-derived cluster: Duroc (brown), Landrace

(pink), Yorkshire White (blue), and Pietrain (purple). Asterisks indicate nominal or genome-wide significance levels as shown in the figure.

## Discussion

The rapid expansion of large human biobanks and livestock reference populations marks a pivotal shift in the dissection of complex traits[7,46]. In both domains, the integration of data from diverse genetic backgrounds is increasingly prioritized to reduce ancestry-related gaps in human genetic studies and to improve the accuracy of joint genetic evaluations in livestock[47-50]. However, this expansion entails a fundamental trade-off, larger sample sizes can increase discovery power, but population stratification and effect-size heterogeneity can confound association tests or attenuate subgroup-dependent signals[29,51,52]. UCALM addresses this challenge by coupling adaptive unsupervised subgroup inference with a layered association framework that integrates heterogeneous association signal, thereby shifting the analytical focus from population labels to underlying genetic effect architectures. By validating UCALM in the UK Biobank and a heterogeneous commercial pig cohort, we demonstrate that the explicit modeling of fine-scale population structure and effect heterogeneity substantially improves discovery power while maintaining rigorous stratification control across species.

A key innovation of UCALM lies in its departure from rigid predefined labels, such as self-reported ethnicity in humans or registered breed in livestock[53]. Instead, UCALM adopts an adaptive density-based clustering on a non-linear genetic manifold, enabling population structure to be inferred directly from genomic data. In modern animal production, extensive crossbreeding and composite-line development generate complex genetic backgrounds that often cannot be adequately represented by conventional breed categorization, paralleling the continuous ancestry gradients observed in admixed human populations. By integrating PCA-UMAP with Auto-HDBSCAN, UCALM effectively resolves fine-scale substructures[17,54], including subtle stratification within European populations and cryptic breeding lines within pig breeds. Moreover, the reassignment procedure based on stable core clusters enables reliable recovery of individuals with missing, incomplete or uncertain annotations, such as crossbred pigs lacking detailed records or admixed humans without reliable self-reported ancestry information. Such individuals are often excluded during conventional quality control, whereas UCALM can retain them for downstream analysis. This maximizes the utilization of available data, increasing effective sample size and improving the representation of genetically diverse individuals, thereby enhancing both the power and generalizability of downstream analyses in human genetics and animal breeding.[46,55,56].

Beyond improving the performance of genetic association, the automated clustering strategy also provides a practical computational advantage. By partitioning a large heterogeneous cohort into smaller and more homogeneous subgroups, UCALM reduces the sample size of each individual-level GWAS, thereby lowering memory requirements and computational burden for downstream association testing. This divide-and-conquer design is particularly useful for very large biobank-scale datasets, where fitting models involving large, dense genomic relationship matrices can impose substantial computational and memory demands[57]. Thus, subgroup-based analysis not only improves the modelling of genetic heterogeneity but also offers a scalable route for association mapping in increasingly large genomic datasets.

Statistically, UCALM addresses the challenge of model selection in meta-analysis by explicitly integrating multiple plausible architectures of allelic effects. Conventional approaches typically favor a single effect model: Fixed-Effects (FE) model is most powerful when effects are homogeneous[25], whereas MR-MEGA is designed to capture ancestry-correlated heterogeneity[29]. However, the causal variants may also exhibit population-specific sparsity[58,59]. The layered meta-analysis strategy in UCALM utilizes the Cauchy combination method to synthesize evidence across homogeneous, ancestry-correlated and population-specific components, thereby retaining sensitivity to whichever architecture best explains the signal. This flexibility is exemplified by the identification of the *MYOT* and *NEK1* loci for backfat thickness in pigs, both of which were obscured in pooled LMM analyses due to its heterogeneous effect size across subgroups. Such findings suggest that missing heritability may fundamentally stem from hidden heterogeneity, particularly for traits and populations in which allele frequencies and linkage disequilibrium differ substantially across genetic backgrounds.

The UCALM framework holds substantial potential for broader applications. In human genetics, the separation of homogeneous and heterogeneous components could enhance the portability of Polygenic Risk Scores (PRS) by prioritizing globally consistent variants across genetic backgrounds[60]. In livestock breeding, UCALM could support cross-population genomic selection by enabling historical datasets from multiple breeds, lines or breeding programs to be integrated more effectively while explicitly accounting for heterogeneity in genetic effects. The resulting association signals could serve as input for downstream genomic prediction models, such as KAML[61] and IFAM[62], that build priors using information from GWAS. By providing a mathematically robust solution to multi-population association mapping, UCALM offers a unified framework to genetic discovery across species. Its ability to distinguish shared, ancestry-correlated and population-specific signals helps ensure that phenotypic associations can be detected and interpreted more rigorously, even in populations with complex genetic structure.

## Methods and Materials

### *The UK Biobank dataset*

The UK Biobank (UKB) is a large-scale prospective cohort study that includes approximately 500,000 participants from diverse ethnic backgrounds, primarily of European ancestry, and offers comprehensive phenotypic records. The genotype data utilized in this investigation were sourced from UKB Data-field 22828, which were imputed by The Wellcome Trust Centre for Human Genetics (WTCHG), Oxford. Quality control (QC) procedures were performed using PLINK software (v2.00)[63]; SNPs were filtered based on: minor allele frequency (MAF) $\geq 0.001$, genotyping rate $\geq 0.95$, and imputation information score $\geq 0.8$. After QC, a total of 487,160 individuals and 14,265,293 SNPs were retained for subsequent analyses. The phenotypic dataset consisted of 24 quantitative traits that were previously analyzed in the fastGWA study[32], including anthropometric, physiological, and biochemical parameters (Supplementary Table 9). For all association models (GLM, UCALM sub-populations, and LDAK-KVIK Step 2), and consistent with the covariate adjustment strategy used in the fastGWA study, we fitted age, $\text{age}^2$, sex, $\text{age} \times \text{sex}$, $\text{age}^2 \times \text{sex}$, and the top 20 principal components as covariates. The sex-specific traits were adjusted for age, $\text{age}^2$ and the top 20 PCs. For LDAK-KVIK, we implemented the recommended two-step protocol to optimize computational efficiency. In Step 1, model parameters were estimated using a prioritized subset of 602,585 independent variants, generated by intersecting the imputed genotype data with HapMap3 variants and performing linkage disequilibrium pruning using PLINK2 (window size 1,000, step size 100, $r^2 > 0.9$)[64]. In Step 2, association statistics were computed for all 14,265,293 retained variants using the full set of covariates. Given the computational burden of fitting standard linear mixed models (LMMs) in large-scale human cohorts and the relatively simple genetic structure of the UKB population, UCALM integrated summary statistics from GLM analyses conducted independently within each inferred subpopulation.

***The Pig dataset***

This study collected genotype data from 85,594 pigs comprising four breeds and three genetic lines. Each pig had 24,439,525 SNPs after genotype imputation. Details of the genotype imputation procedure and the results are provided in the Supplementary Notes (Supplementary Fig. 14). Quality control (QC) was performed on the imputed genotypes using PLINK software (v2.00)[63], applying filters for a minor allele frequency (MAF) ≥ 0.01 and a genotype call rate ≥ 0.95. After QC, the final dataset included 85,594 individuals and 19,713,426 high-confidence SNPs for downstream analyses. Phenotypic records were obtained for four economic traits comprising backfat thickness adjusted to 100 kg (BF, $n = 75{,}538$), age adjusted to 100 kg (AGE, $n = 75{,}981$), feed conversion ratio (FCR, $n = 19{,}822$), and loin muscle area adjusted to 100 kg (LMA, $n = 51{,}785$). Each GWAS was conducted using individuals with phenotypic records for the corresponding trait. Detailed information on the samples and phenotypic measurements is provided in Supplementary Table 10. For the pig cohort, the sample size was computationally tractable for mixed-model analysis, whereas the complex genetic background required explicit control of relatedness and population structure; therefore, pooled general linear model (GLM) and linear mixed model (LMM) analyses were conducted in the full cohort, while UCALM used LMM-based association analyses within each inferred subpopulation. All GWAS models included sex, year-season, and the top 5 principal components as covariates.

Independent association signals were defined by partitioning the genome into 100-kb windows. Applying a genome-wide significance threshold of $P < 2.54 \times 10^{-9}$, we identified the lead SNP exhibiting the lowest $P$-value within each significant window. To quantify the additional power provided by our method, UCALM-detected signals were classified as novel if the corresponding 100-kb window lacked significant association signals in the standard LMM analysis; conversely, signals located in windows that also contained significant LMM hits were classified as common. Functional annotation was performed using BEDTools (v2.31.1) and the *Sus scrofa 11.1.104* reference genome[65]. Lead SNPs were mapped to genes within a 100-kb window and prioritized hierarchically based on their genomic context: exonic, intronic, proximal and distal.

### *Adaptive automated clustering and stratification*

We implemented an unsupervised dimensionality-reduction and density-based clustering procedure to infer fine-scale population structure directly from high-density genotype data. First, a genomic relationship matrix (GRM) was constructed from genome-wide markers and subjected to eigen decomposition to obtain the leading $k$ principal components, with $k$ set to 3 by default and also adjustable by the user. In this study, $k = 10$ was used for downstream clustering analyses. To improve computational efficiency for large cohorts, we linked to the *Spectra* library

(https://spectralib.org), enabling rapid estimation of the top $k$ eigenvectors without full matrix decomposition. In the second stage, the resulting PCs were used as input for UMAP to obtain a two-dimensional embedding of genetic relationships among individuals. To account for differences in variance explained across PCs, each retained eigenvector was variance-weighted according to its associated eigenvalue before UMAP, so components explaining more variance contributed more strongly to the distance calculations. Specifically, let $u_i = (u_{i1}, u_{i2}, \ldots u_{ik})^T$ denote the coordinates of individual $i$ on the leading $k$ eigenvectors of the GRM, where $u_{ij}$ is the coordinate on the $j_{th}$ retained eigenvector $(j = 1, 2, \ldots, k)$. The corresponding eigenvalues are $\lambda_1 \geq \lambda_2 \geq \cdots \geq \lambda_k$, with $\bar{\lambda} = k^{-1} \sum_{j=1}^{k} \lambda_j$. The variance-weighted PC vector of individual $i$, denoted $p_i$, was defined as:

$$p_i = \frac{1}{\sqrt{\bar{\lambda}}} \left( \sqrt{\lambda_1} u_{i1}, \sqrt{\lambda_2} u_{i2}, \ldots, \sqrt{\lambda_\mathrm{k}} u_{ik} \right)^T \tag{1}$$

UMAP generates a low-dimensional representation:

$$z_i = UMAP(p_i; n_{nn}, d_{min}) \tag{2}$$

where $z_i$ is the two-dimensional embedding, $n_{nn}$ is the number of nearest neighbors, and $d_{min}$ is the UMAP minimum-distance parameter, which controls the compactness of local neighborhoods in the low-dimensional embedding and was set to 0.1 by default.

Subsequently, to delineate population substructures through a fully automated procedure, without requiring user-defined thresholds or manual intervention, we applied an optimized Hierarchical Density-Based Spatial Clustering of Applications with Noise (HDBSCAN) algorithm. For two samples $x_i$ and $x_j$, the mutual reachability distance is defined as:

$$d_{\mathrm{mreach},k}(x_\mathrm{i}, x_j) = \max\{ \mathrm{Core}_k(x_i), \mathrm{Core}_k(x_j), d(x_i, x_j)\} \tag{3}$$

where $d(x_i, x_j)$ is the Euclidean distance in the reduced space, and $\mathrm{Core}_k(x_i)$ is the distance to the $k_{th}$ nearest neighbor, where $k$ corresponds to $\mathrm{minPts}$, ensuring robustness against local density variation. A minimum spanning tree (MST) is constructed using Prim's algorithm based on these distances and recursively partitioned to generate a hierarchical cluster tree[66].

The granularity of clustering is critically controlled by the minimum number of points parameter $\mathrm{minPts}$. In the program, $\mathrm{minPts}$ was set directly according to the UMAP nearest-neighbor parameter:

$$\mathrm{minPts} = n_{nn} \tag{4}$$

If $n_{nn}$ was not specified, it was set adaptively to the smaller value between 300 and 5% of the sample size, thereby defining the local neighborhood size used by UMAP. For UMAP neighbor search, the software selected the nearest-neighbor index according to the dimensionality of the retained PC feature space. When more than 50 PC features were retained, a Hierarchical Navigable Small World (HNSW)[67] index was used because graph-based approximate nearest-neighbor search is more scalable in higher-dimensional spaces, although it may introduce a small approximation error. Otherwise, a Vantage-Point tree (VP-tree)[68] with Euclidean distance was used, because metric-tree search is efficient and deterministic in lower-dimensional spaces but becomes less effective as dimensionality increases. The cutoff of 50 retained PC features was used as a practical implementation threshold, reflecting the reduced efficiency of exact metric-tree searches in higher-dimensional spaces and the better scalability of graph-based approximate nearest-neighbor search under such settings. After UMAP embedding, HDBSCAN was applied to the two-dimensional coordinates, with k-dimensional tree (KD-tree)[69] search used to accelerate core-distance and nearest-neighbor calculations in the UMAP space.

Standard density-based methods often discard admixed or peripheral individuals as noise. To maximize data utilization while monitoring stratification risk, we implemented an iterative label propagation strategy based on the nearest assigned neighbor. Initial HDBSCAN clustering was performed in the UMAP space, whereas noise reassignment was performed in the variance-weighted PCA space. This design preserves the sensitivity of UMAP-based clustering to local structure while assigning ambiguous samples using distances derived directly from the retained PCs. Let $C$ be the set of currently assigned individuals, initialized with individuals belonging to the stable core clusters, and let $N$ be the set of unassigned noise points. For each unassigned noise point $i$, each $j \in C$ represents a candidate assigned individual, and the index of its nearest assigned individual $a(i)$ is defined as:

$$a(i) = arg\min_{j \in C} \left\| p_i - p_j \right\|_2^2 \quad (5)$$

Noise points are ranked by their proximity to stable clusters and reassigned iteratively. At each step, the noise point with the smallest distance is assigned to the cluster of its nearest assigned individual, added to $C$, and then used to update the nearest assigned individual for the remaining noise points. Crucially, we introduce a stability indicator $S_i$ for every individual: samples originally clustered by HDBSCAN are labeled as Core ($S_i = 0$), whereas reassigned samples are flagged as Ambiguity ($S_i = 1$). This annotation allows downstream analyses to distinguish between robust cluster members and ambiguously assigned individuals.

***Sub-population association testing***

Following population partitioning, Genome-Wide Association Studies (GWAS) were conducted separately within each inferred population. In principle, the framework can accommodate any GWAS model. Here, we employed the standard linear mixed model (LMM) used in the pig dataset as an example, the LMM model can be described as:

$$y = Xb + Z\mu + e \quad (6)$$

where $y$ is a vector of phenotypic records for $n$ individuals; $b$ is a vector of the estimated effects of the fixed covariates with the corresponding coefficient matrix $X$; $\mu$ is a vector of individual genetic values assumed to follow $\mu \sim N(0, G\sigma_g^2)$ with the corresponding variance-covariance matrix $G$ that is also defined as genomic relationship matrix (GRM); $e$ is the vector of residual effects, assumed to follow $e \sim$ $N(0, I\sigma_e^2)$, where $I$ is an identity matrix; the $\sigma_g^2$ and $\sigma_e^2$ are the estimated genetic variance and residual variance, respectively.

***Layered meta-analysis framework***

To capture a broad spectrum of genetic architectures, from homogeneous to ancestry-correlated and population-specific effects, we proposed a two-layer meta-analysis framework. In the first layer, we first employed an optimized Meta-Regression of Multi-Ancestry Genetic Association (MR-MEGA) approach to capture ancestry-correlated effects[29]. Specifically, the genome was partitioned into fixed-length windows (for example, 1 Mb), and a pairwise Euclidean distance matrix $D = [d_{kk'}]$ was constructed for each window using allele frequencies of all SNPs within that region:

$$d_{kk'} = \sqrt{\sum_j \left(p_{kj} - p_{k'j}\right)^2} \tag{7}$$

Eigen decomposition was then performed on $D$ to extract the top $t$ principal axes of genetic variation $(v_{kt})$, where $t$ is determined adaptively based on the number of populations $(N_{pops})$ as follows:

$$t = \begin{cases} 3 & \text{if } N_{pops} > 3 \\ N_{pops} - 1 & \text{if } N_{pops} \leq 3 \end{cases} \tag{8}$$

The effect size of variant $j$ in population $k$, denoted as $\beta_{kj}$, is modeled as:

$$\beta_{kj} = \alpha_j + \sum_{i=1}^{t} \gamma_{ij} v_{ki} + \varepsilon_{kj} \tag{9}$$

where $\gamma_{ij}$ represents the change in effect size along the $i_{th}$ ancestry axis. The model was fitted using Weighted Least Squares (WLS), with weights defined as $\omega_{kj} = 1/se_{kj}^2$. We then test the null hypothesis that the variant has no effect in any population (i.e., intercept and all slopes are zero), yielding the Meta-Regression *P*-value, $P_{MR}$.

Next, to capture signals with common effect sizes across populations, we calculated the standard Inverse Variance Weighted (IVM) Fixed-Effects statistic[25]:

$$\beta_{FE} = \frac{\sum_k \omega_{kj}\beta_{kj}}{\sum_k \omega_{kj}}, \quad Z_{FE} = \frac{\beta_{FE}}{\sqrt{1/\sum_k \omega_{kj}}} \tag{10}$$

The two-tailed $P$-value derived from $Z_{FE}$ is denoted as $P_{FE}$.

To construct an omnibus test that is robust to signal sparsity or population-specific heterogeneity, as well as statistical correlation, we utilized the Aggregated Cauchy Association Test (ACAT)[34,35] to combined the $P$-value for SNP $j$ across all $k$ sub-populations $(P_{1j}, \ldots, P_{kj})$, with weights assigned in proportion to the sample size of each sub-population to reflect their relative contribution:

$$T_{PS} = \sum_{i=1}^{k} \omega_i \tan\left(\left(0.5 - P_{ij}\right)\pi\right) \tag{11}$$

where $\omega_i = N_i / N_{total}$ represents the weight of the $i_{th}$ sub-population, $N_i$ is the sample size of the $i_{th}$ sub-population, and $N_{total}$ is the total sample size across all populations. The resulting $P$-value $P_{PS}$ is derived from a standard Cauchy distribution and is specifically designed to capture population-specific signals, which typically arise when genetic effects are present in only one or a few populations.

In the second layer of meta-analysis, we integrated the three complementary components, representing homogeneous effects $P_{FE}$, ancestry-correlated heterogeneous effects $P_{MR}$, and population-specific effects $P_{Ps}$, into the final association statistic:

$$T_{UCALM} = \sum_{m=1}^{M} \omega_m \tan\left(\left(0.5 - P_m\right)\pi\right) \tag{12}$$

where $M$ is the number of association components, $P_m$ denotes the $P$-value from the $m$-th component, and $\omega_m$ is the corresponding weight, with $\sum_{m=1}^{M} \omega_m = 1$. In this study, $M = 3$, corresponding to $P_{FE}$, $P_{MR}$, and $P_{PS}$, and equal weights were used ($\omega_m = 1/3$), giving each architecture of genetic effects the same prior contribution. The final UCALM $P$-value was calculated as:

$$P_{UCALM} = 0.5 - \frac{\arctan(T_{UCALM})}{\pi} \tag{13}$$

By applying the Cauchy combination, UCALM yields an omnibus test that remains valid under arbitrary dependency among $P_{FE}$, $P_{MR}$, and $P_{PS}$. Its heavy-tailed transformation provides high sensitivity to strong evidence from any single component while still integrating association evidence across the three complementary signal architectures.

***Simulating phenotypes***

To extensively evaluate the statistical power and Type I error control of our proposed framework, we performed simulations based on real genotype data from the UK biobank (UKB) dataset. We randomly selected 50,000 unrelated individuals comprising four distinct ancestral populations: European (EUR, $n = 35{,}000$), African (AFR, $n = 7{,}000$), South Asian (SAS, $n = 6{,}600$), and East Asian (EAS, $n = 1{,}400$). To enable a rigorous assessment of both statistic power and false positive rates within the unified simulation framework, we partitioned the genome according to physical position. For each chromosome, the centromeric midpoint was identified, and the first half of the chromosome was used to select causal variants for power evaluation, whereas the second half was used to select null variants for Type I error assessment. A 4Mb buffer region was introduced between these two segments to minimize linkage disequilibrium (LD) spillover and ensure statistical independence between causal and null signals.

Phenotypes were generated under a standard additive genetic model defined as $y = X\beta + e$, where $y$ denotes the vector of simulated phenotypes, $X$ represents the genotype matrix of causal variants, $\beta$ is the vector of effect sizes, and $e$ is the residual error. The variance of the residuals was determined by the target heritability $(h^2)$ according to the formula $\sigma_e^2 = \sigma_g^2(1 - h^2)/h^2$, where $\sigma_g^2$ is the genetic variance. We simulated quantitative traits across varying heritability levels $(h^2 \in \{0.2, 0.5, 0.8\})$, performing 20 independent replications for each experimental setting to ensure statistical robustness.

We designed four distinct genetic architectures to cover the major classes of association signals expected in multi-population cohorts. First, homogeneous effects were simulated using 1,000 causal variants with relatively stable allele frequencies across populations (maximum allele-frequency difference < 0.10), with constant effect sizes across populations drawn independently from a normal distribution with mean 0 and variance 1. Second, ancestry-correlated heterogeneity was simulated using 1,000 variants whose effect sizes varied linearly with population differentiation. Specifically, a base effect was scaled by population-specific gradient factors of 1, -0.25, 0.25 and 0 for EUR, AFR, SAS and EAS, respectively. Third, population-specific effects were simulated using 100 variants showing marked allele frequency divergence, defined as loci for which the allele frequency in EAS exceeded that in every other population by at least 0.20. For these variants, genetic effects were set to be non-zero only in the EAS group and zero in all other groups. Finally, a mixed architecture was simulated by combining homogeneous and ancestry-correlated variants, yielding a total of 1,000 causal SNPs, to evaluate simultaneous detection of shared and stratified effects. All simulated phenotypes were standardized prior to analysis to ensure consistent scale across methods.

***Assessing the statistical power and false positive rate***

To systematically benchmark the performance of our proposed UCALM, we compared it with the LMM and three representative meta-analysis approaches, including FE, RE, and MR-MEGA, across 20 independent simulation replicates. Statistical power was evaluated using two complementary metrics derived from simulated causal SNPs: the mean $\chi^2$ test statistic to quantify overall signal strength, and the empirical power, defined as the count of causal variants reaching the rigorous Bonferroni threshold $(p < 5.56 \times 10^{-9})$. Simultaneously, to ensure that improved sensitivity did not compromise Type I error control, we evaluated the genomic inflation of null SNPs from preserved genomic regions by computing the mean $\chi^2$ statistic, and assessed robustness across significance levels by calculating the empirical False Positive Rate (FPR) at four nominal thresholds $(0.05, 5 \times 10^{-3}, 5 \times 10^{-4}, and\ 5 \times 10^{-5})$.

**Data availability**

The individual-level genotype and phenotype data of humans are sourced from the UK biobank (http://www.ukbiobank.ac.uk), it is accessible for authors under Application Number 97563 (https://www.ukbiobank.ac.uk/enable-your-research/apply-for-access). The pig dataset analyzed in this study has been deposited in the Science Data Bank (https://doi.org/10.57760/sciencedb.44216) and will be publicly available on August 4, 2028. Before public release, requests for access to the data should be directed to the corresponding author and will be considered on a reasonable-request basis, subject to approval by the data owner.

## Code availability

UCALM has been incorporated into our MVP software, a C++ successor to rMVP[70], and is freely available to the public. The MVP software and detailed user manual can be accessed freely at: https://mvp-cpp.github.io/tutorial/ucalm/.

LDAK.KVIK is available at: https://www.ldak-kvik.com/.

Plink (v2.00) is available at: https://www.cog-genomics.org/plink/2.0/.

Bedtools (v2.31.1) is available at: https://github.com/arq5x/bedtools2.

## Competing interests

The authors declare no competing interests.

## Acknowledgments

This work was supported by the National Natural Science Foundation of China [32494801] to X.L.L., the Fundamental and Interdisciplinary Disciplines Breakthrough Plan of the Ministry of Education of China [JYB2025XDXM704] to X.L.L., L.L.Y., and Y.H.F., the Yazhouwan National Laboratory Fundamental Research Project [2310SH01] to Z.S.T. and S.H.Z., the earmarked fund for CARS [CARS-35] to S.H.Z., X.Y.L, and X.L.L.

## Author contributions

X.L.L., L.L.Y., and X.Y.L. conceived the study. L.L.Y., and X.L.L. designed the experiments. X.S., Y.L., and H.H.Z. developed the software tools under the assistance and guidance from L.L.Y.; X.S., and Z.S.T. performed the data collection, simulations and analyses under the assistance and guidance from Z.Y.L., H.J.Z., H.H.Z., Y.F.L., D.Y., Y.W., Y.Q., X.X., Y.H.F., S.H.Z., X.Y.L., L.L.Y., and X.L.L.; X.S., and L.L.Y. wrote the raw paper, Z.S.T., S.H.Z., X.Y.L., L.L.Y., and X.L.L. modified the paper. All authors reviewed and approved the final manuscript.

## References

1. MacArthur, J. *et al.* The new NHGRI-EBI Catalog of published genome-wide association studies (GWAS Catalog). *Nucleic acids research* **45**, D896-D901 (2017).
2. Visscher, P.M. *et al.* 10 years of GWAS discovery: biology, function, and translation. *The American Journal of Human Genetics* **101**, 5-22 (2017).
3. Wang, Q., Tang, J., Han, B. & Huang, X. Advances in genome-wide association studies of complex traits in rice. *Theoretical and Applied Genetics* **133**, 1415-1425 (2020).
4. Xu, Z. *et al.* Integrating large-scale meta-GWAS and PigGTEx resources to decipher the genetic basis of 232 complex traits in pigs. *National Science Review* **12**, nwaf048 (2025).
5. Yan, G. *et al.* An imputed whole-genome sequence-based GWAS approach pinpoints causal mutations for complex traits in a specific swine population. *Science China Life Sciences* **65**, 781-794 (2022).
6. Bycroft, C. *et al.* The UK Biobank resource with deep phenotyping and genomic data. *Nature* **562**, 203-209 (2018).
7. Consortium, U.B.W.-G.S. Whole-genome sequencing of 490,640 UK Biobank participants. *Nature* **645**, 692 (2025).
8. Du, H. *et al.* The 1000 Chinese Indigenous Pig Genomes Project provides insights into the genomic architecture of pigs. *Nature Communications* **15**, 10137 (2024).
9. Feng, Y.-C.A. *et al.* Taiwan Biobank: A rich biomedical research database of the Taiwanese population. *Cell genomics* **2**(2022).
10. Hayes, B.J. & Daetwyler, H.D. 1000 bull genomes project to map simple and complex genetic traits in cattle: applications and outcomes. *Annual review of animal biosciences* **7**, 89-102 (2019).
11. Kurki, M.I. *et al.* FinnGen provides genetic insights from a well-phenotyped isolated population. *Nature* **613**, 508-518 (2023).
12. Martin, A.R. *et al.* Clinical use of current polygenic risk scores may exacerbate health disparities. *Nature genetics* **51**, 584-591 (2019).
13. Price, A.L., Zaitlen, N.A., Reich, D. & Patterson, N. New approaches to population stratification in genome-wide association studies. *Nature reviews genetics* **11**, 459-463 (2010).
14. Yu, J. *et al.* A unified mixed-model method for association mapping that accounts for multiple levels of relatedness. *Nature genetics* **38**, 203-208 (2006).

15. Price, A.L. *et al.* Principal components analysis corrects for stratification in genome-wide association studies. *Nature genetics* **38**, 904-909 (2006).
16. Özkaraca, M.İ., Agung, M., Navarro, P. & Tenesa, A. Divide and conquer approach for genome-wide association studies. *Genetics* **229**, iyaf019 (2025).
17. McInnes, L., Healy, J. & Melville, J. Umap: Uniform manifold approximation and projection for dimension reduction. *arXiv preprint arXiv:1802.03426* (2018).
18. Bengio, Y., Goodfellow, I. & Courville, A. *Deep learning*, (MIT press Cambridge, MA, USA, 2017).
19. Clarke, R. *et al.* The properties of high-dimensional data spaces: implications for exploring gene and protein expression data. *Nature reviews cancer* **8**, 37-49 (2008).
20. Alexander, D.H., Novembre, J. & Lange, K. Fast model-based estimation of ancestry in unrelated individuals. *Genome research* **19**, 1655 (2009).
21. Pritchard, J.K., Stephens, M. & Donnelly, P. Inference of population structure using multilocus genotype data. *Genetics* **155**, 945-959 (2000).
22. Raj, A., Stephens, M. & Pritchard, J.K. fastSTRUCTURE: variational inference of population structure in large SNP data sets. *Genetics* **197**, 573-589 (2014).
23. Bulik-Sullivan, B.K. *et al.* LD Score regression distinguishes confounding from polygenicity in genome-wide association studies. *Nature genetics* **47**, 291-295 (2015).
24. Loh, P.-R. *et al.* Efficient Bayesian mixed-model analysis increases association power in large cohorts. *Nature genetics* **47**, 284-290 (2015).
25. Cochran, W.G. The combination of estimates from different experiments. *Biometrics* **10**, 101-129 (1954).
26. Morris, A.P. Transethnic meta-analysis of genomewide association studies. *Genetic epidemiology* **35**, 809-822 (2011).
27. Willer, C.J., Li, Y. & Abecasis, G.R. METAL: fast and efficient meta-analysis of genomewide association scans. *Bioinformatics* **26**, 2190-2191 (2010).
28. Han, B. & Eskin, E. Random-effects model aimed at discovering associations in meta-analysis of genome-wide association studies. *The American Journal of Human Genetics* **88**, 586-598 (2011).
29. Mägi, R. *et al.* Trans-ethnic meta-regression of genome-wide association studies accounting for ancestry increases power for discovery and improves fine-mapping resolution. *Human molecular genetics* **26**, 3639-3650 (2017).
30. Campello, R.J., Moulavi, D. & Sander, J. Density-based clustering based on hierarchical density estimates. in *Pacific-Asia conference on knowledge discovery and data mining* 160-172 (Springer, 2013).
31. Hof, J.P. & Speed, D. LDAK-KVIK performs fast and powerful mixed-model association analysis of quantitative and binary phenotypes. *Nature Genetics* **57**, 2116-2123 (2025).
32. Jiang, L. *et al.* A resource-efficient tool for mixed model association analysis of large-scale data. *Nature genetics* **51**, 1749-1755 (2019).

33. Mbatchou, J. *et al.* Computationally efficient whole-genome regression for quantitative and binary traits. *Nature genetics* **53**, 1097-1103 (2021).
34. Liu, Y. *et al.* ACAT: a fast and powerful p value combination method for rare-variant analysis in sequencing studies. *The American Journal of Human Genetics* **104**, 410-421 (2019).
35. Liu, Y. & Xie, J. Cauchy combination test: a powerful test with analytic p-value calculation under arbitrary dependency structures. *Journal of the American Statistical Association* **115**, 393-402 (2020).
36. Yang, J. *et al.* Genomic inflation factors under polygenic inheritance. *European journal of human genetics* **19**, 807-812 (2011).
37. Yengo, L. *et al.* Meta-analysis of genome-wide association studies for height and body mass index in~ 700000 individuals of European ancestry. *Human molecular genetics* **27**, 3641-3649 (2018).
38. Israeli, H. *et al.* Structure reveals the activation mechanism of the MC4 receptor to initiate satiation signaling. *Science* **372**, 808-814 (2021).
39. Duan, Y. *et al.* VRTN is required for the development of thoracic vertebrae in mammals. *International Journal of Biological Sciences* **14**, 667 (2018).
40. Van Laere, A.-S. *et al.* A regulatory mutation in IGF2 causes a major QTL effect on muscle growth in the pig. *Nature* **425**, 832-836 (2003).
41. Sebag, J.A., Zhang, C., Hinkle, P.M., Bradshaw, A.M. & Cone, R.D. Developmental control of the melanocortin-4 receptor by MRAP2 proteins in zebrafish. *Science* **341**, 278-281 (2013).
42. Chen, X.-Y. *et al.* Brain-selective kinase 2 (BRSK2) phosphorylation on PCTAIRE1 negatively regulates glucose-stimulated insulin secretion in pancreatic β-cells. *Journal of Biological Chemistry* **287**, 30368-30375 (2012).
43. Mead, T.J. ADAMTS6: Emerging roles in cardiovascular, musculoskeletal and cancer biology. *Frontiers in Molecular Biosciences* **9**, 1023511 (2022).
44. Davis, R.L., Weintraub, H. & Lassar, A.B. Expression of a single transfected cDNA converts fibroblasts to myoblasts. *Cell* **51**, 987-1000 (1987).
45. White, M.C. & Quarmby, L.M. The NIMA-family kinase, Nek1 affects the stability of centrosomes and ciliogenesis. *BMC cell biology* **9**, 29 (2008).
46. Wojcik, G.L. *et al.* Genetic analyses of diverse populations improves discovery for complex traits. *Nature* **570**, 514-518 (2019).
47. Hong, E. *et al.* Effect of breed composition in genomic prediction using crossbred pig reference population. *Journal of Animal Science and Technology* **67**, 56 (2025).
48. Wang, H. *et al.* Improving Genetic Risk Prediction of CAD in Chinese by Multi-ancestry and Multi-trait GWAS Integration. *Genomics, Proteomics & Bioinformatics*, qzag038 (2026).
49. Wang, Y. *et al.* Theoretical and empirical quantification of the accuracy of polygenic scores in ancestry divergent populations. *Nature communications* **11**, 3865 (2020).

50. Zhou, J. *et al.* Evaluating the performance of genomic selection on purebred population by incorporating crossbred data in pigs. *Journal of Integrative Agriculture* **23**, 639-648 (2024).
51. Peterson, R.E. *et al.* Genome-wide association studies in ancestrally diverse populations: opportunities, methods, pitfalls, and recommendations. *Cell* **179**, 589-603 (2019).
52. Privé, F. *et al.* Portability of 245 polygenic scores when derived from the UK Biobank and applied to 9 ancestry groups from the same cohort. *The American Journal of Human Genetics* **109**, 12-23 (2022).
53. Mathieson, I. & Scally, A. What is ancestry? *PLoS genetics* **16**, e1008624 (2020).
54. Diaz-Papkovich, A., Anderson-Trocmé, L., Ben-Eghan, C. & Gravel, S. UMAP reveals cryptic population structure and phenotype heterogeneity in large genomic cohorts. *PLoS genetics* **15**, e1008432 (2019).
55. Karaman, E., Su, G., Croue, I. & Lund, M.S. Genomic prediction using a reference population of multiple pure breeds and admixed individuals. *Genetics Selection Evolution* **53**, 46 (2021).
56. Zhang, H. *et al.* A new method for multiancestry polygenic prediction improves performance across diverse populations. *Nature genetics* **55**, 1757-1768 (2023).
57. Yin, L. *et al.* HIBLUP: an integration of statistical models on the BLUP framework for efficient genetic evaluation using big genomic data. *Nucleic acids research* **51**, 3501-3512 (2023).
58. Rubin, C.-J. *et al.* Strong signatures of selection in the domestic pig genome. *Proceedings of the National Academy of Sciences* **109**, 19529-19536 (2012).
59. Tropf, F.C. *et al.* Hidden heritability due to heterogeneity across seven populations. *Nature human behaviour* **1**, 757-765 (2017).
60. Ding, Y. *et al.* Polygenic scoring accuracy varies across the genetic ancestry continuum. *Nature* **618**, 774-781 (2023).
61. Yin, L. *et al.* KAML: improving genomic prediction accuracy of complex traits using machine learning determined parameters. *Genome biology* **21**, 146 (2020).
62. Tang, Z. *et al.* Improving genomic prediction accuracy of complex traits by integrating massive types of functional annotation information. *Nature Communications* (2026).
63. Chang, C.C. *et al.* Second-generation PLINK: rising to the challenge of larger and richer datasets. *Gigascience* **4**, s13742-015-0047-8 (2015).
64. Consortium, I.H. Integrating common and rare genetic variation in diverse human populations. *Nature* **467**, 52 (2010).
65. Quinlan, A.R. & Hall, I.M. BEDTools: a flexible suite of utilities for comparing genomic features. *Bioinformatics* **26**, 841-842 (2010).
66. Prim, R.C. Shortest connection networks and some generalizations. *The Bell System Technical Journal* **36**, 1389-1401 (1957).

67. Malkov, Y.A. & Yashunin, D.A. Efficient and robust approximate nearest neighbor search using hierarchical navigable small world graphs. *IEEE transactions on pattern analysis and machine intelligence* **42**, 824-836 (2018).
68. Yianilos, P. Data structures and algorithms for nearest neighbor search in general metric spaces. *Proceedings of the fourth annual ACM-SIAM…* (1993).
69. Bentley, J.L. Multidimensional binary search trees used for associative searching. *Communications of the ACM* **18**, 509-517 (1975).
70. Yin, L. *et al.* rMVP: a memory-efficient, visualization-enhanced, and parallel-accelerated tool for genome-wide association study. *Genomics, proteomics & bioinformatics* **19**, 619-628 (2021).

# Supplementary Information

## A unified unsupervised framework for genome-wide association studies in heterogeneous populations

Xiong Shen[1,2†], Zhenshuang Tang[2†], Yong Liao[1], Zhiyuan Lin[1], Huajun Zhou[1], Haohao Zhang[3], Yangfan Liu[1], Dong Yin[1], Yue Wang[1], Yuan Quan[7], Zhuqing Zheng[8], Xiong Xiong[1,4], Yuhua Fu[1,5], Shuhong Zhao[1,2,4,5,6], Xinyun Li[1,4,5,6*], Lilin Yin[1,4,5,6*], Xiaolei Liu[1,4,5,6*]

[1] *Key Laboratory of Pig Digital and Intelligent Breeding, Ministry of Agriculture and Rural Affairs & College of Animal Science and Technology, Huazhong Agricultural University, Wuhan 430070, China;*

[2] *Yazhouwan National Laboratory, Sanya 572024, China;*

[3] *School of Computer Science and Technology, Wuhan University of Technology, Wuhan 430070, China;*

[4] *Hubei Hongshan Laboratory, Wuhan 430070, China;*

[5] *Key Laboratory of Agricultural Animal Genetics, Breeding and Reproduction, Ministry of Education, Wuhan 430070, China;*

[6] *Frontiers Science Center for Animal Breeding and Sustainable Production, Ministry of Education, Wuhan 430070, China;*

[7] *College of Informatics, Huazhong Agricultural University, Wuhan 430070, China;*

[8] *Institute of Food Nutrition and Health, Jingchu University of Technology, Jingmen, Hubei 448000, China;*

[†] *These authors contributed equally to this work;*

[*] *Corresponding authors: Xinyun Li (xyli@mail.hzau.edu.cn), Lilin Yin (ylilin@mail.hzau.edu.cn), Xiaolei Liu (xiaoleiliu@mail.hzau.edu.cn).*

# Supplementary Note

### The UK Biobank dataset and processing

The UK Biobank (UKB) imputation release contains approximately 96 million markers before study-specific filtering[1]. Genomic coordinates were based on GRCh37[2]. All UKB analyses in this study were restricted to autosomal SNPs; sex-chromosome markers were excluded before clustering and association testing. No additional sample-level filtering was applied beyond the analysis set described in the Methods section. Covariate information was taken from UKB Data-field 21022 for age, Data-field 31 for sex and Data-field 22009 for genetic principal components. These fields were used to construct the covariates described in the Methods section. Supplementary Table 9 provides the trait information for the 24 UKB quantitative phenotypes, including trait abbreviation, UKB data-field ID, number of records, phenotype category and full trait name. Phenotypes were analyzed on their original scales. Self-reported ethnicity was taken from UKB Data-field 21000. These labels were not used for UCALM clustering. They were used only to describe the inferred UKB clusters and to compare data-driven clusters with available sample labels. For the LDAK-KVIK comparison, Step 1 variants were selected from the UKB genotype set after intersection with HapMap3 variants[3] using PLINK v2.00 with --indep-pairwise 1000 100 0.9[4]. Starting from 1,169,312 HapMap3 variants, LD pruning removed 566,727 variants and retained 602,585 variants for Step 1 model fitting. This follows the LDAK-KVIK recommendation for dense datasets, where Step 1 can be run on a subset of predictors to reduce runtime, whereas all retained genome-wide variants are still used in Step 2 association testing[5].

**The Pig Dataset and processing**

Tail tissue samples were collected from each pig during routine tail docking in piglets. This sampling was integrated into standard farm management to reduce additional handling. Immediately after collection, tail tissues were preserved in 75% ethanol and stored at -20°C to maintain DNA integrity. All preserved samples were transported to Yingzi Gene Technology Co., Ltd. (Wuhan, China) for genomic analysis. Genomic DNA was extracted from tail tissues using standard protocols. High throughput genotyping was performed using the pig 80K functional variants genotyping array. The genotype data with 187,000 SNPs were imputed using the reference panel and imputation tool downloaded from the AGIDB website[6]. Variants with imputation accuracy $R^2 > 0.7$ were retained.

Imputation accuracy was assessed by genotype masking in 20-fold cross-validation. Biallelic SNPs shared between the chip genotype dataset and the phased reference panel were divided into 20 mutually exclusive folds of approximately equal size. In each round, one-fold was masked from the chip genotype dataset and imputed with Beagle v5.5[7], while the remaining chip variants were kept as observed genotypes. The imputed genotypes at masked loci were then compared with the corresponding observed genotypes. Imputation accuracy was measured by consistency (CR) and correlation ($R^2$). CR was defined as the proportion of correctly imputed hard call genotypes among all non-missing observed genotypes. $R^2$ was calculated as the squared Pearson correlation between imputed and observed genotype dosages, with observed genotypes coded as 0, 1 and 2 and converted to minor allele dosage when needed. Both metrics were summarized at the individual and SNP levels. The results across the 20 cross-validation folds are shown in Supplementary Fig. 14. Both individual level and SNP level summaries showed high CR and $R^2$, supporting high imputation accuracy for individuals and markers in the pig dataset. Supplementary Table 10 provides the phenotype information for the four pig traits, including trait abbreviation, number of records, trait category and full trait description. Supplementary Data 2 provides the final pig cluster composition. In the pig labels, the three-letter code denotes company, breed and genetic line.

**The Simulated Dataset**

The simulation analyses used the 50,000-individual UKB panel described in the Methods section. Quality control procedures were performed using PLINK v2.00, SNPs were filtered based on: minor allele frequency (MAF) ≥ 0.01, genotyping rate ≥ 0.95. After QC, a total of 50,000 individuals and 8,996,037 SNPs were retained for subsequent analyses. For LDAK-KVIK, we used the same recommended two-step protocol as in the UKB real trait analysis to improve computational efficiency[5]. In Step 1, model parameters were estimated using a prioritized subset of 643,087 independent variants, generated by intersecting the imputed genotype data with HapMap3 variants and performing linkage disequilibrium pruning using PLINK v2.00 with --indep-pairwise 1000 100 0.9[4]. In Step 2, association statistics were computed for all 8,996,037 retained variants using the full set of covariates.

**The Simulation Evaluation Metrics**

Simulation results were summarized using metrics calculated separately for causal and null SNP sets. Let $C$ denote the set of simulated causal SNPs and $N$ denote the set of null SNPs. For each method and replicate, let $P_s$ be the association P-value for SNP $s$, and let $\chi_s^2$ be the corresponding one degree of freedom chi-square statistic. For causal SNPs, association strength was quantified by the mean chi-square statistic:

$$\overline{\chi_C^2} = \frac{1}{|C|}\sum_{s \in C} \chi_s^2$$

The number of detected causal SNPs was evaluated using the Bonferroni threshold:

$$K = \sum_{s \in C} I(P_s < 0.05/M)$$

where $M$ is the number of SNPs tested in the simulation analysis and $I(\cdot)$ is the indicator function. This count was used as the empirical power metric.

For null SNPs, inflation of null test statistics was assessed using the mean chi-square statistic:

$$\overline{\chi_N^2} = \frac{1}{|N|} \sum_{s \in N} \chi_s^2$$

The empirical type I error rate, reported as the false positive rate (FPR) at significance level $\alpha$, was calculated as:

$$FPR(\alpha) = \frac{1}{|N|} \sum_{s \in N} I(P_s < \alpha)$$

The evaluated thresholds were $0.05$, $5 \times 10^{-3}$, $5 \times 10^{-4}$ and $5 \times 10^{-5}$. For reference, the binomial standard error under the nominal false positive rate $\alpha$ was calculated across 20 simulation replicates as:

$$SE(\alpha) = \sqrt{\frac{\alpha(1-\alpha)}{20|N|}}.$$

**Cauchy combination test in UCALM**

UCALM used the Cauchy combination test to integrate dependent component *P*-values. In the first layer, subgroup *P*-values were combined to detect population specific signals, with weights proportional to the number of samples contributing to each subgroup test. In the second layer, the three-component $P$-values for homogeneous, ancestry correlated and population specific signals were combined with equal weights. These component $P$-values are calculated from overlapping subgroup summary statistics and are therefore not independent. The Cauchy combination test is suitable for this setting because it combines $P$-values without estimating the covariance matrix among component tests[8,9].

For input $P$-values $P_1, \cdots, P_K$ and non-negative weights $\omega_1, \cdots, \omega_K$, the implemented statistic was:

$$T = \frac{\sum_{k=1}^{K} \omega_k \tan\{(0.5 - P_k)\pi\}}{\sum_{k=1}^{K} \omega_k}$$

The combined P-value was then calculated as:

$$P_{Cauchy} = 0.5 - \frac{\arctan(T)}{\pi}$$

For the second-layer combination of the homogeneous, ancestry-correlated and population-specific components, this combined $P$-value corresponds to the final UCALM $P$-value, denoted as $P_{UCALM}$.

In UCALM, the Cauchy transformation gives greater influence to components with stronger association signals while retaining all components in the same weighted statistic. This property helps preserve sensitivity when the association signal is mainly captured by the homogeneous, ancestry correlated or population specific component. If multiple components show association signals, their P-values enter the same weighted Cauchy statistic after transformation and jointly influence $P_{UCALM}$.

For numerical stability, the implementation used tail approximations in extreme cases. When $P_k < 10^{-15}$, the Cauchy transform was evaluated using:

$$\tan\{(0.5 - P_k)\pi\} = \cot(\pi P_k) \approx \frac{1}{\pi P_k}$$

which avoids numerical overflow in the tangent function. When the combined statistic was very large $(T > 10^{15})$, the final *P*-value was evaluated using:

$$P_{Cauchy} \approx \frac{1}{\pi T}$$

These tail approximations were used only to improve numerical stability for extremely small *P*-values or very large Cauchy statistics. They do not change the definition of the Cauchy combination statistic.

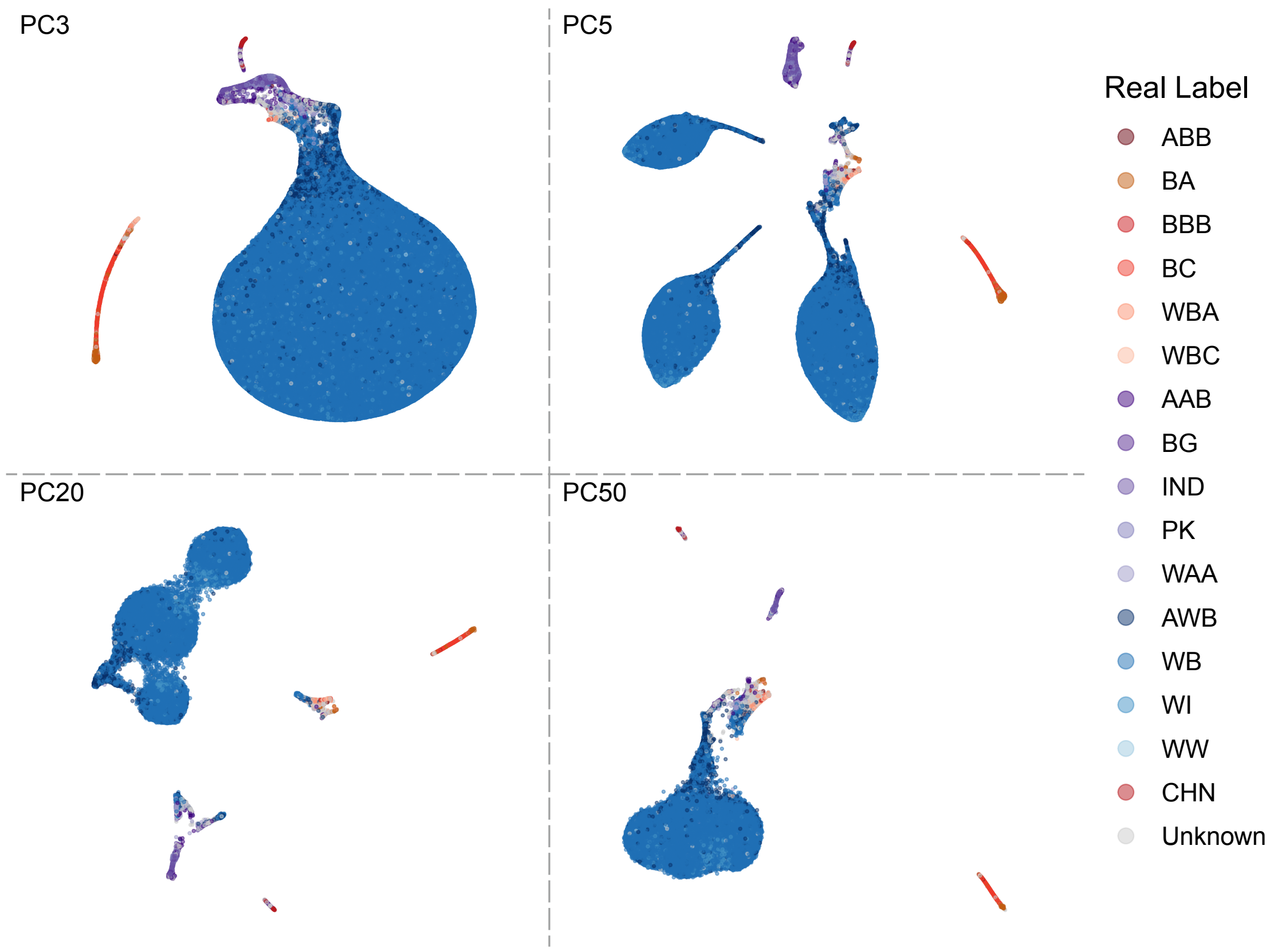

**Supplementary Fig. 1.** UMAP projections of UKB individuals generated using different numbers of principal components (PCs). Panels correspond to UMAP projections generated from 3, 5, 20 and 50 PCs, with points colored by self-reported ethnicity labels. Abbreviations: ABB, Any other Black background; BA, Black African; BBB, Black or Black British; BC, Black Caribbean; WBA, White and Black African; WBC, White and Black Caribbean; AAB, Any other Asian background; BG, Bangladeshi; IND, Indian; PK, Pakistani; WAA, White and Asian; AWB, Any other white background; WB, White British; WI, White Irish; WW, White; and CHN, Chinese. Unknown includes samples labelled as AMB, Any other mixed background; DNK, Do not know; OEG, Other ethnic group; ASB, Asian or Asian British; and NA.

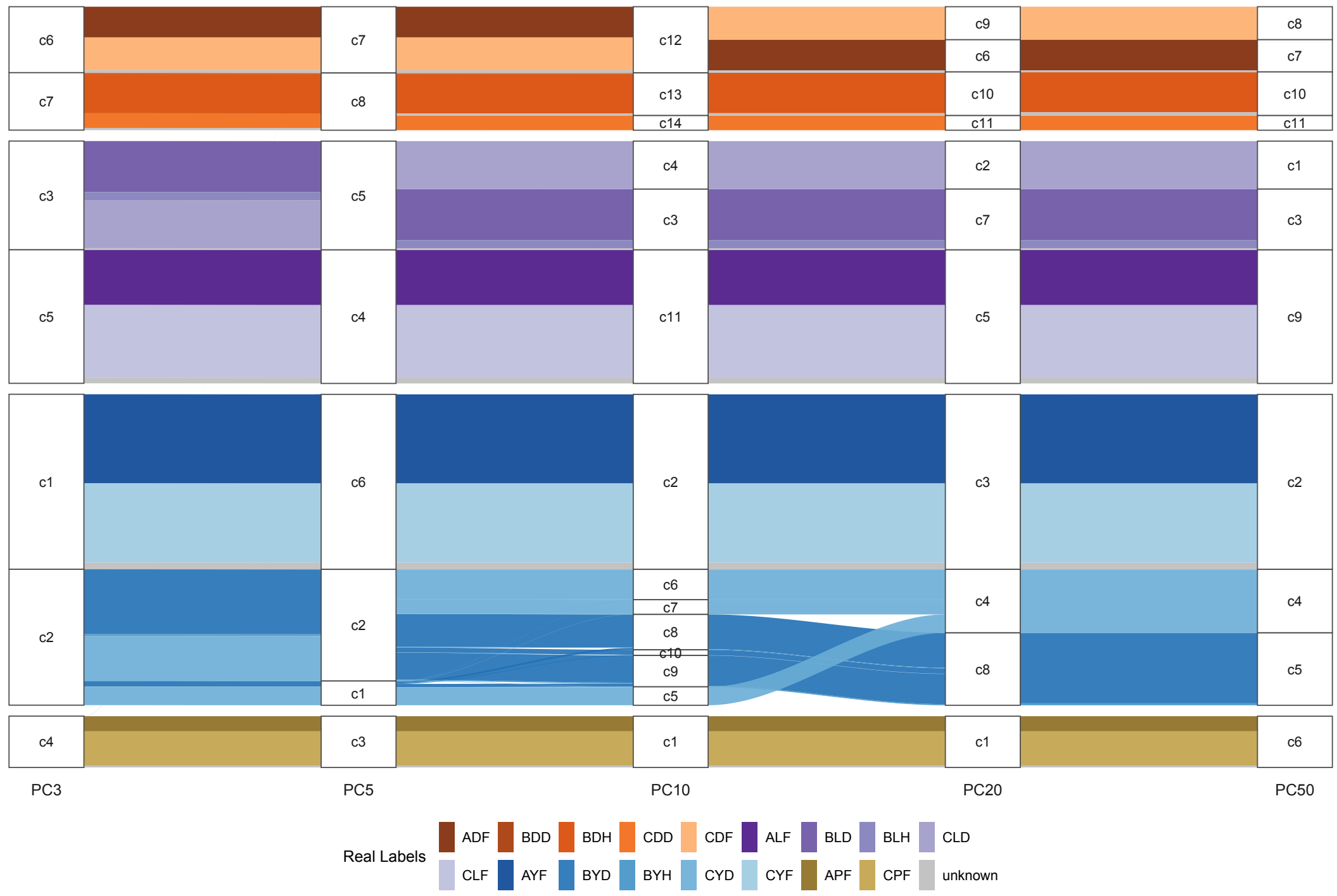

**Supplementary Fig. 2.** Alluvial plot of pig cluster assignments across different numbers of principal components (PCs). Columns show cluster assignments obtained from UMAP projections generated with 3, 5, 10, 20 and 50 PCs. The height of each cluster block and the width of each stream are proportional to the number of individuals. Streams are colored by reference pig population labels, encoded as three-letter codes. In each code, the first letter indicates company (A, B or C), the second letter indicates breed (D, Duroc; L, Landrace; Y, Yorkshire; P, Pietrain), and the third letter indicates line (F, French; D, Danish; H, Chinese). Samples lacking complete records are labelled as unknown and shown in grey.

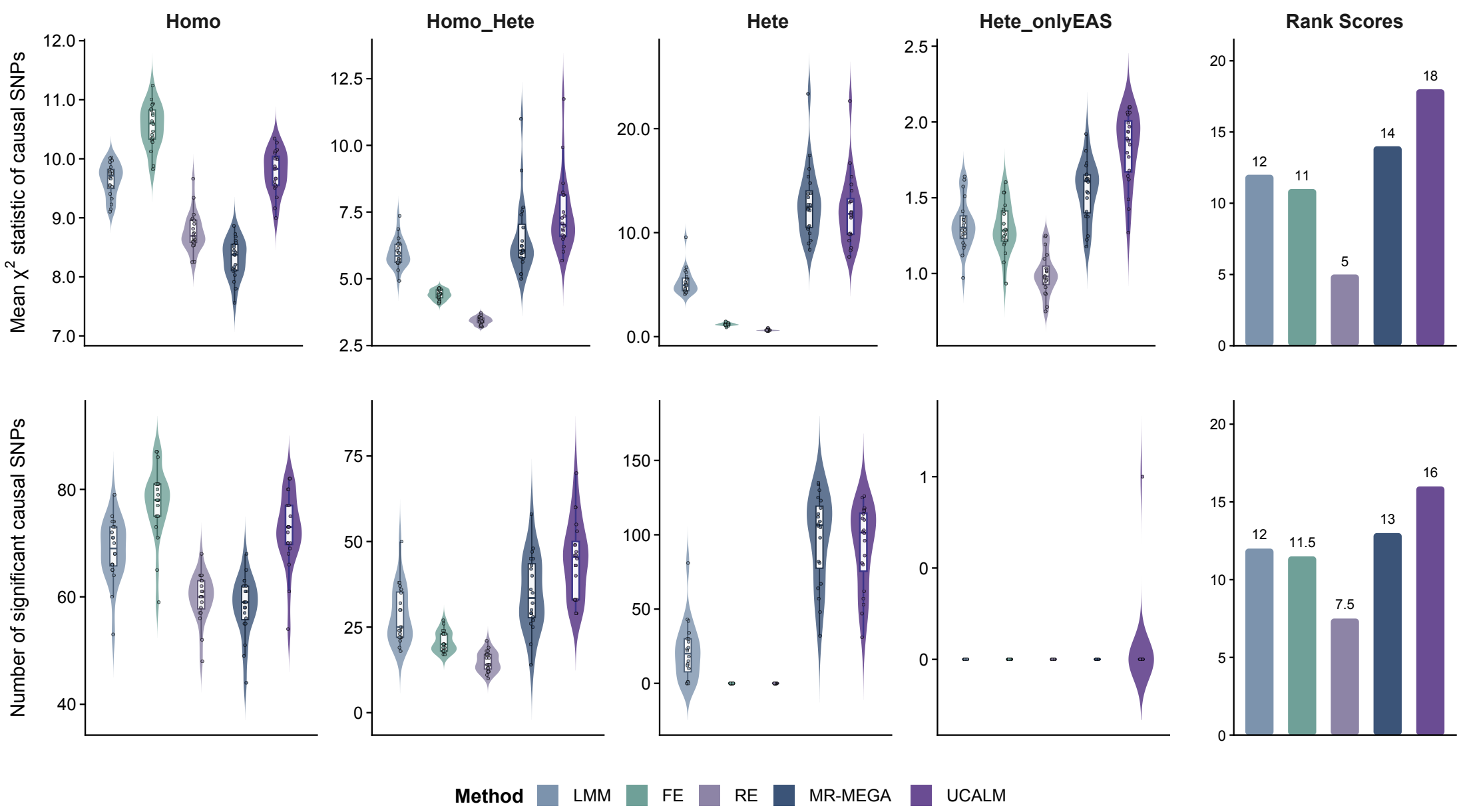

**Supplementary Fig. 3.** Association signal strength and number of significant causal SNPs across four simulated genetic architectures at $h^2 = 0.2$. Performance of LMM, FE, RE, MR-MEGA and UCALM was evaluated across 20 independent simulation replicates. Columns show homogeneous effects (Homo), mixed homogeneous and heterogeneous effects (Homo_Hete), ancestry-correlated heterogeneous effects (Hete), and EAS-specific effects (Hete_onlyEAS). The top row shows the mean $\chi^2$ statistic for causal SNPs, reflecting association signal strength across replicates. The bottom row shows the number of significant causal SNPs across replicates. Rightmost bar plots show the corresponding method rank scores for each metric. When methods showed identical performance, they were assigned the average of the corresponding ranked scores. In violin plots, the internal boxplot marks the median and interquartile range (IQR), and whiskers extend to $1.5 \times$ IQR.

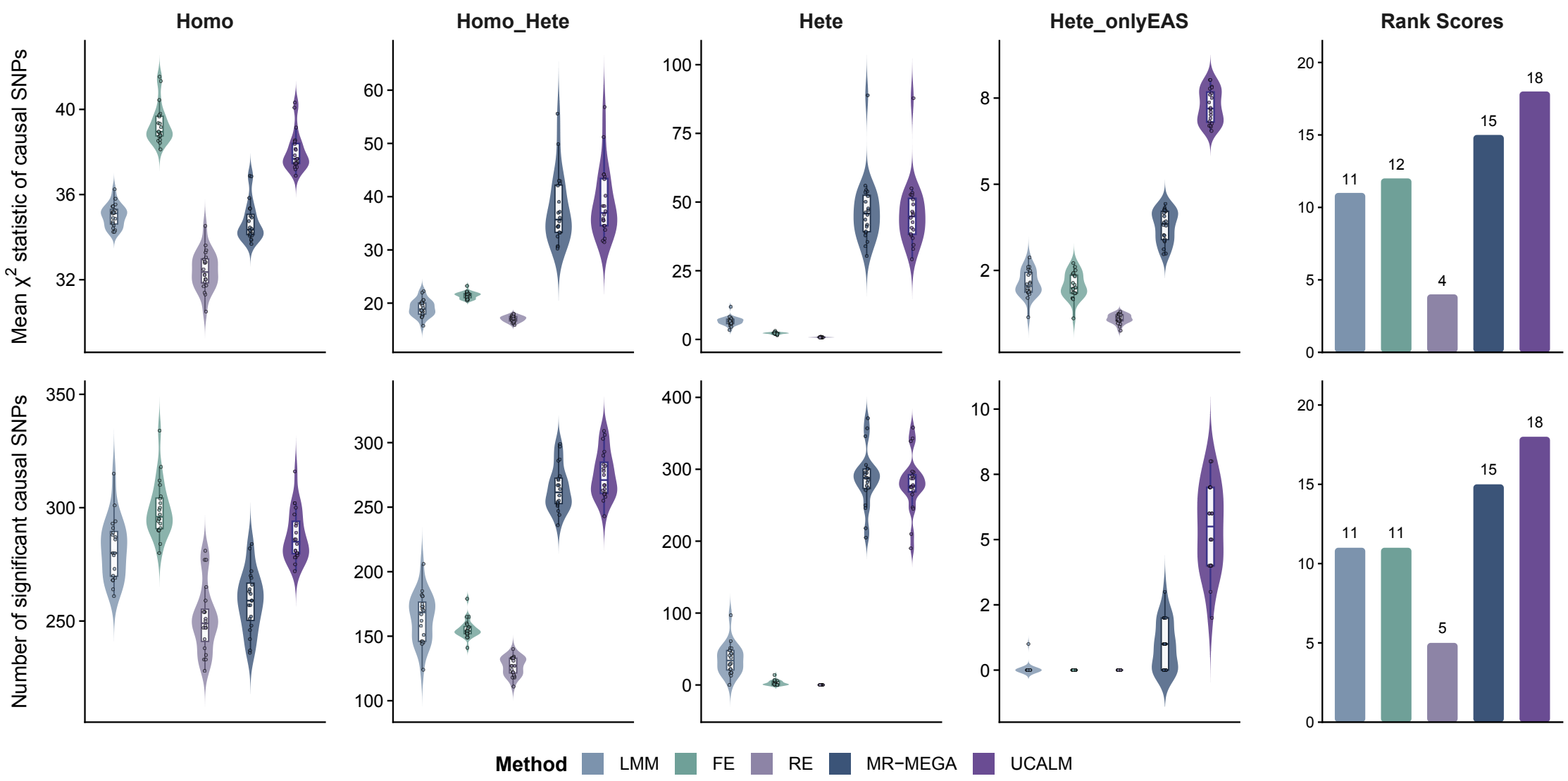

**Supplementary Fig. 4.** Association signal strength and number of significant causal SNPs across four simulated genetic architectures at $h^2 = 0.8$. Performance of LMM, FE, RE, MR-MEGA and UCALM was evaluated across 20 independent simulation replicates. Columns show homogeneous effects (Homo), mixed homogeneous and heterogeneous effects (Homo_Hete), ancestry-correlated heterogeneous effects (Hete), and EAS-specific effects (Hete_onlyEAS). The top row shows the mean $\chi^2$ statistic for causal SNPs, reflecting association signal strength across replicates. The bottom row shows the number of significant causal SNPs across replicates. Rightmost bar plots show the corresponding method rank scores for each metric. When methods showed identical performance, they were assigned the average of the corresponding ranked scores. In violin plots, the internal boxplot marks the median and interquartile range (IQR), and whiskers extend to $1.5 \times$ IQR.

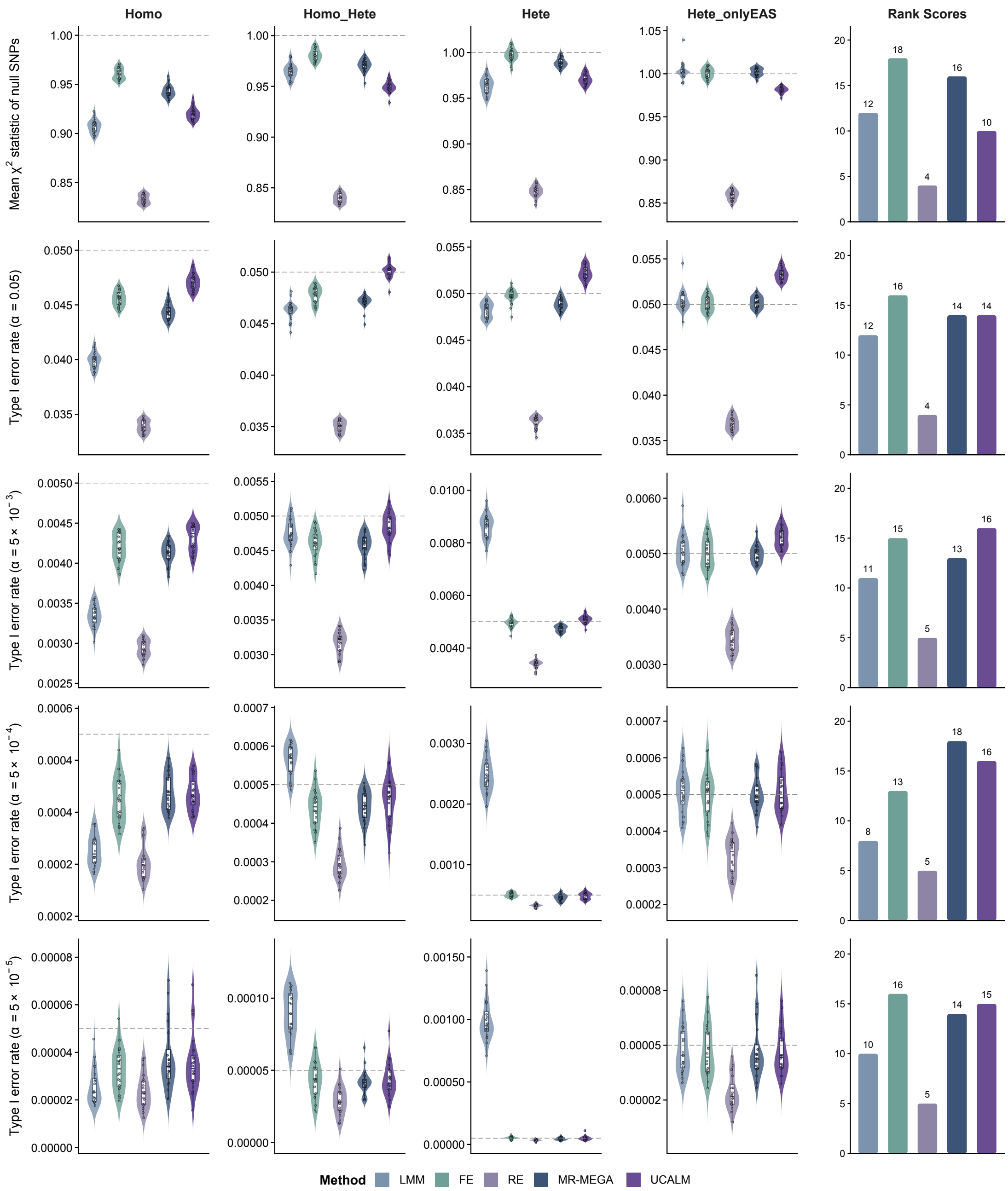

**Supplementary Fig. 5.** Type I error control across four simulated genetic architectures at $h^2 = 0.2$. Performance of LMM, FE, RE, MR-MEGA and UCALM was evaluated across 20 independent simulation replicates. Columns show homogeneous effects (Homo), mixed homogeneous and heterogeneous effects (Homo_Hete), ancestry-correlated heterogeneous effects (Hete), and EAS-specific effects (Hete_onlyEAS). The top row shows the mean $\chi^2$ statistic for null SNPs. The remaining rows show empirical type I error rates at $\alpha = 0.05$, $5 \times 10^{-3}$, $5 \times 10^{-4}$, and $5 \times 10^{-5}$. Rightmost bar plots show the corresponding method rank scores for each metric. When methods showed identical performance, they were assigned the average of the corresponding ranked scores. In violin plots, the internal boxplot marks the median and interquartile range (IQR), and whiskers extend to $1.5 \times$ IQR. Dashed grey lines indicate expected values under the null hypothesis: 1.00 for the mean $\chi^2$ statistic and the nominal $\alpha$ level for type I error rates.

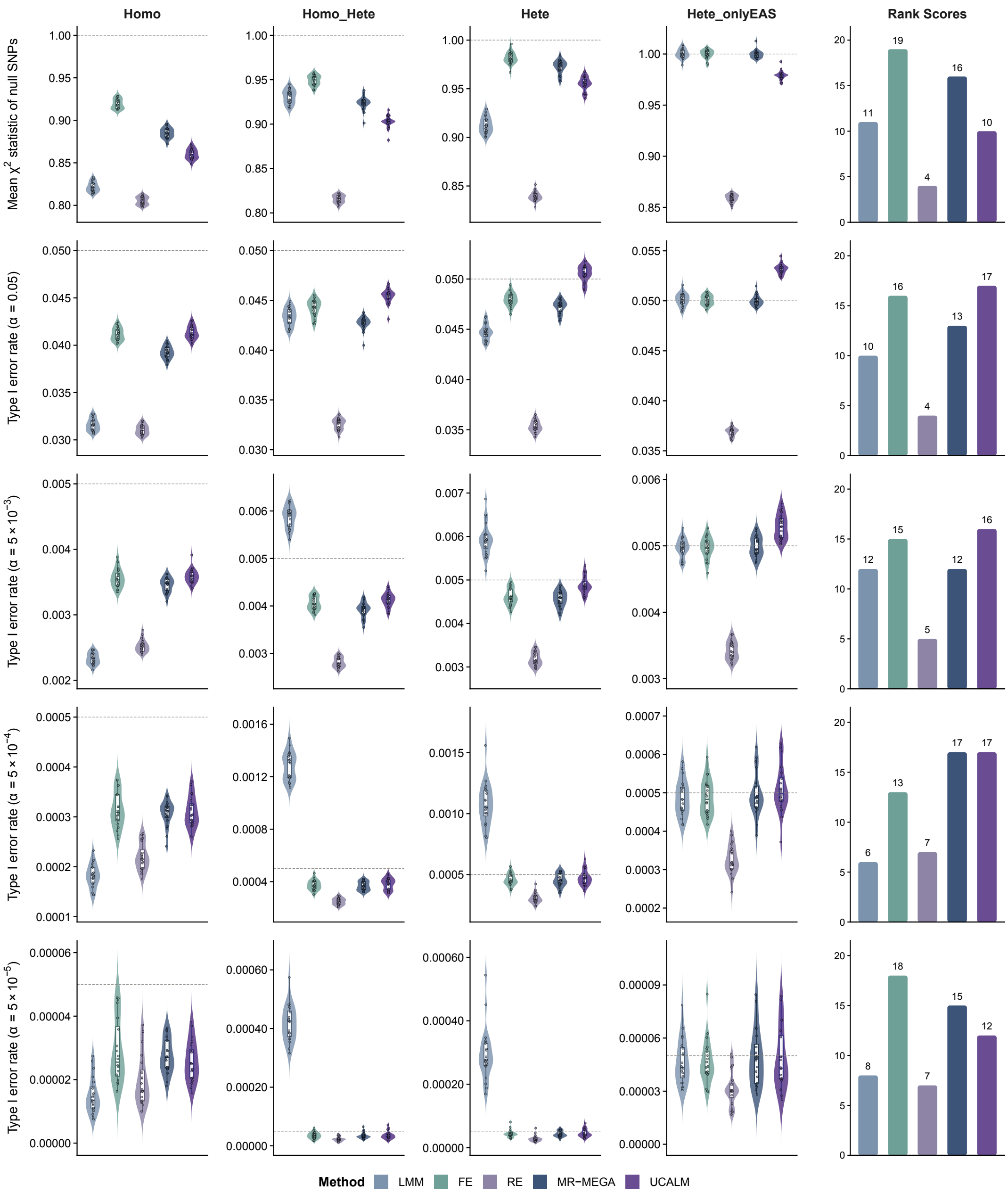

**Supplementary Fig. 6.** Type I error control across four simulated genetic architectures at $h^2 = 0.5$. Performance of LMM, FE, RE, MR-MEGA and UCALM was evaluated across 20 independent simulation replicates. Columns show homogeneous effects (Homo), mixed homogeneous and heterogeneous effects (Homo_Hete), ancestry-correlated heterogeneous effects (Hete), and EAS-specific effects (Hete_onlyEAS). The top row shows the mean $\chi^2$ statistic for null SNPs. The remaining rows show empirical type I error rates at $\alpha = 0.05$, $5 \times 10^{-3}$, $5 \times 10^{-4}$, and $5 \times 10^{-5}$. Rightmost bar plots show the corresponding method rank scores for each metric. When methods showed identical performance, they were assigned the average of the corresponding ranked scores. In violin plots, the internal boxplot marks the median and interquartile range (IQR), and whiskers extend to $1.5 \times$ IQR. Dashed grey lines indicate expected values under the null hypothesis: 1.00 for the mean $\chi^2$ statistic and the nominal $\alpha$ level for type I error rates.

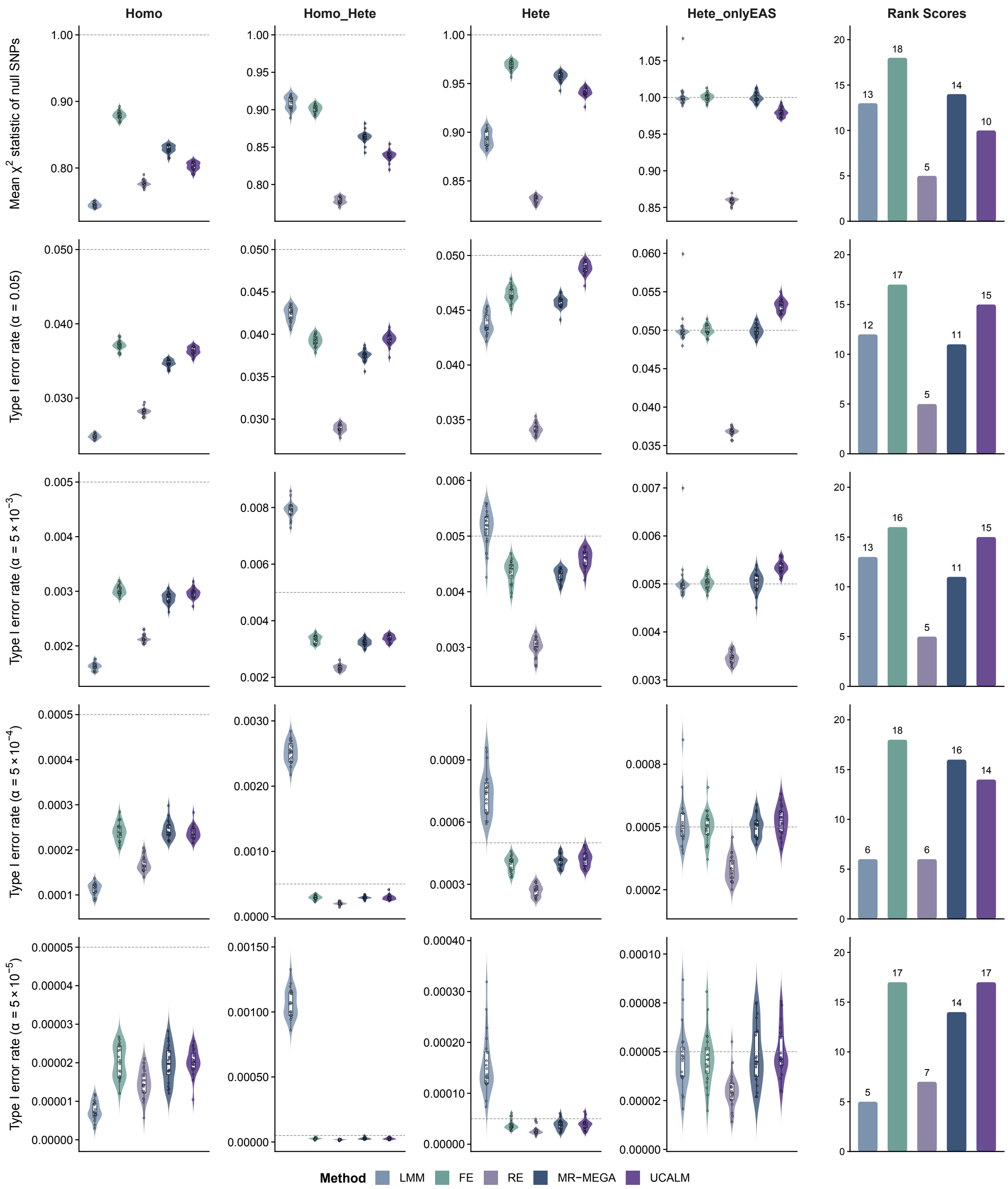

**Supplementary Fig. 7.** Type I error control across four simulated genetic architectures at $h^2 = 0.8$. Performance of LMM, FE, RE, MR-MEGA and UCALM was evaluated across 20 independent simulation replicates. Columns show homogeneous effects (Homo), mixed homogeneous and heterogeneous effects (Homo_Hete), ancestry-correlated heterogeneous effects (Hete), and EAS-specific effects (Hete_onlyEAS). The top row shows the mean $\chi^2$ statistic for null SNPs. The remaining rows show empirical type I error rates at $\alpha = 0.05$, $5 \times 10^{-3}$, $5 \times 10^{-4}$, and $5 \times 10^{-5}$. Rightmost bar plots show the corresponding method rank scores for each metric. When methods showed identical performance, they were assigned the average of the corresponding ranked scores. In violin plots, the internal boxplot marks the median and interquartile range (IQR), and whiskers extend to $1.5 \times$ IQR. Dashed grey lines indicate expected values under the null hypothesis: 1.00 for the mean $\chi^2$ statistic and the nominal $\alpha$ level for type I error rates.

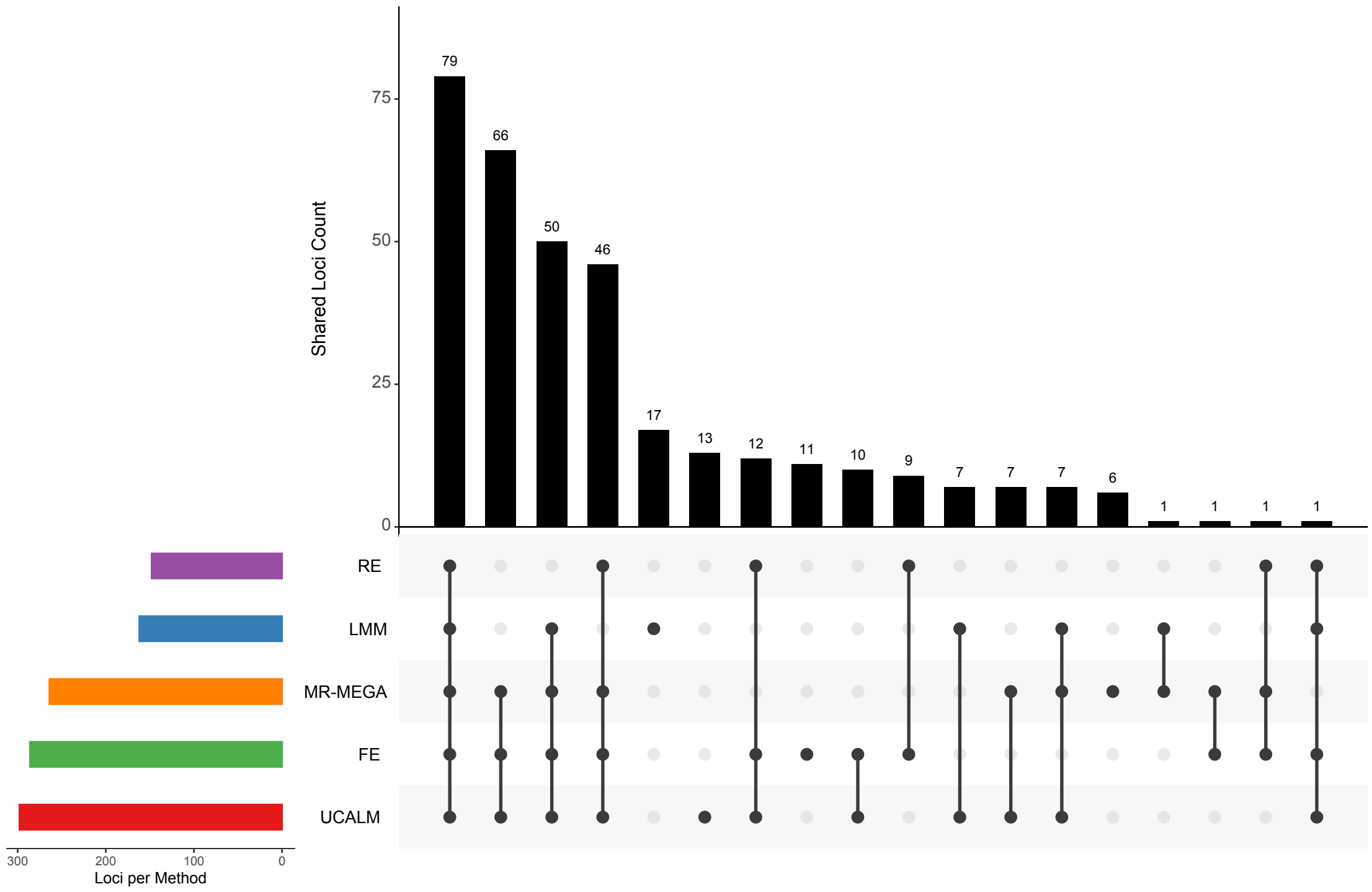

**Supplementary Fig. 8.** UpSet plot of significant loci for the BF trait across five association strategies. The plot shows the overlap among significant 100 kb genomic windows identified by RE, LMM, MR-MEGA, FE and UCALM. Vertical bars indicate the number of loci in each method intersection, and the connected points below each bar indicate the methods included in that intersection. Horizontal bars show the total number of significant loci detected by each method. Methods are ordered by total locus count, and intersections are ordered by decreasing intersection size.

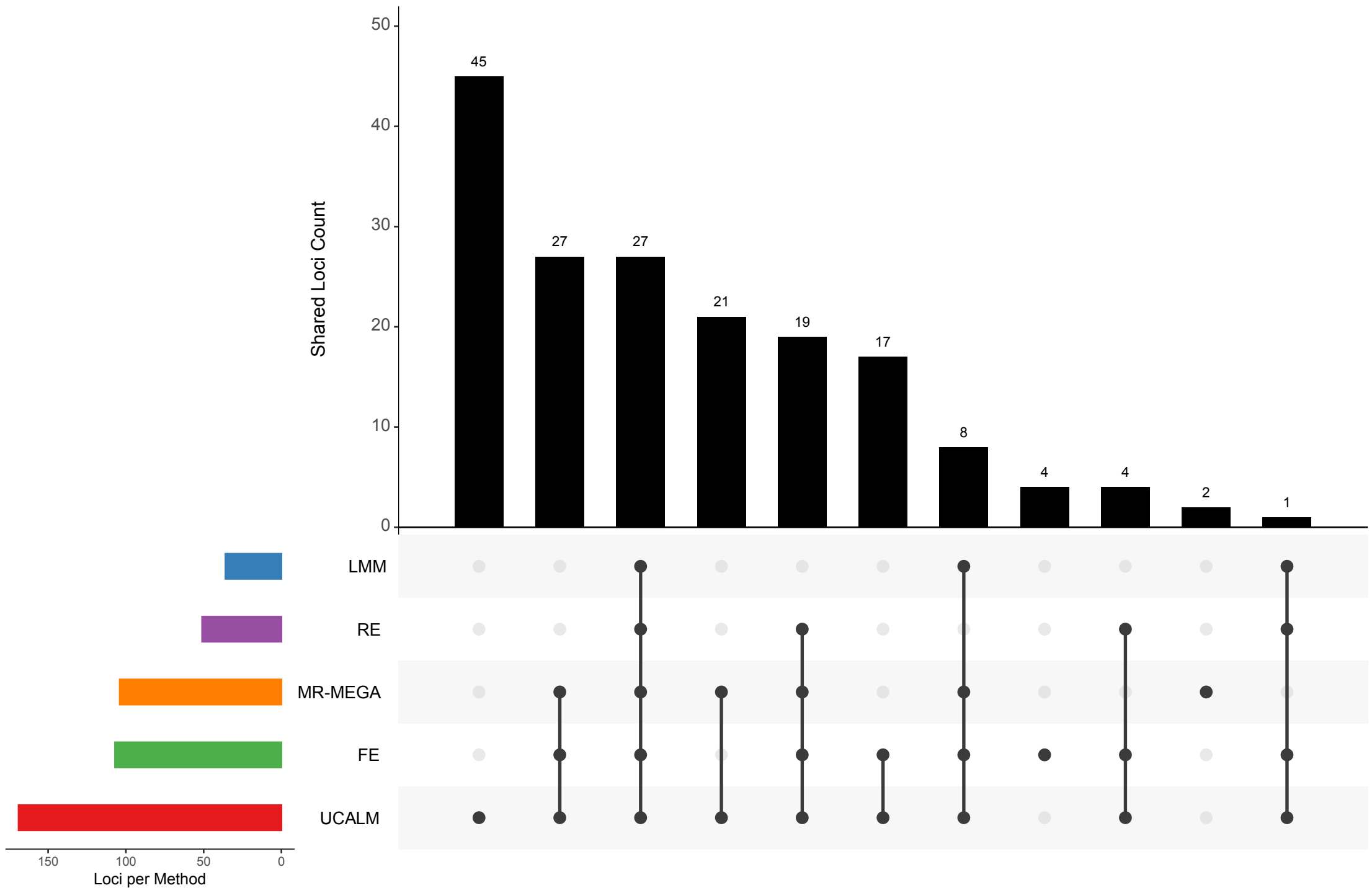

**Supplementary Fig. 9.** UpSet plot of significant loci for the AGE trait across five association strategies. The plot shows the overlap among significant 100 kb genomic windows identified by RE, LMM, MR-MEGA, FE and UCALM. Vertical bars indicate the number of loci in each method intersection, and the connected points below each bar indicate the methods included in that intersection. Horizontal bars show the total number of significant loci detected by each method. Methods are ordered by total locus count, and intersections are ordered by decreasing intersection size.

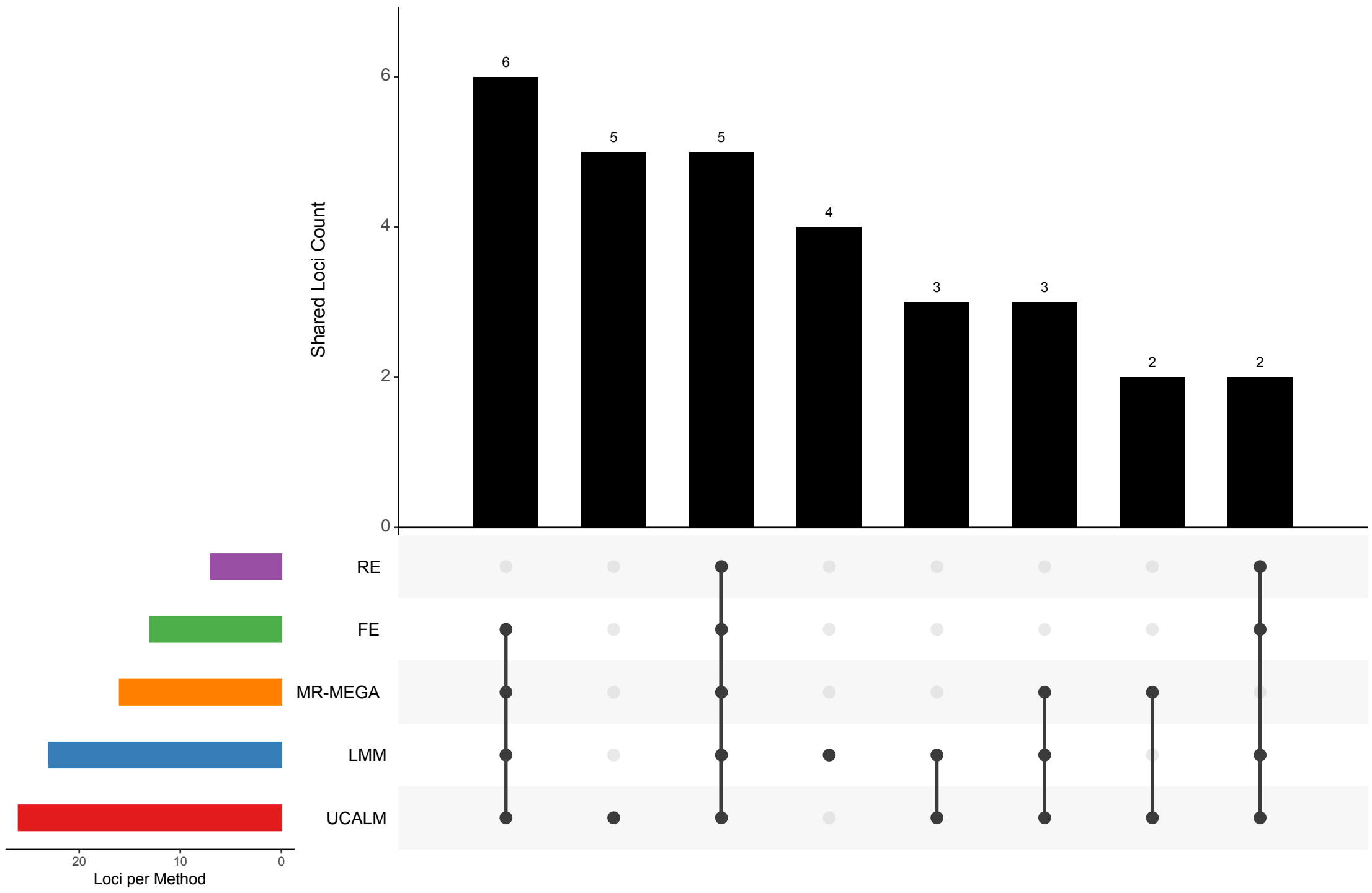

**Supplementary Fig. 10.** UpSet plot of significant loci for the FCR trait across five association strategies. The plot shows the overlap among significant 100 kb genomic windows identified by RE, LMM, MR-MEGA, FE and UCALM. Vertical bars indicate the number of loci in each method intersection, and the connected points below each bar indicate the methods included in that intersection. Horizontal bars show the total number of significant loci detected by each method. Methods are ordered by total locus count, and intersections are ordered by decreasing intersection size.

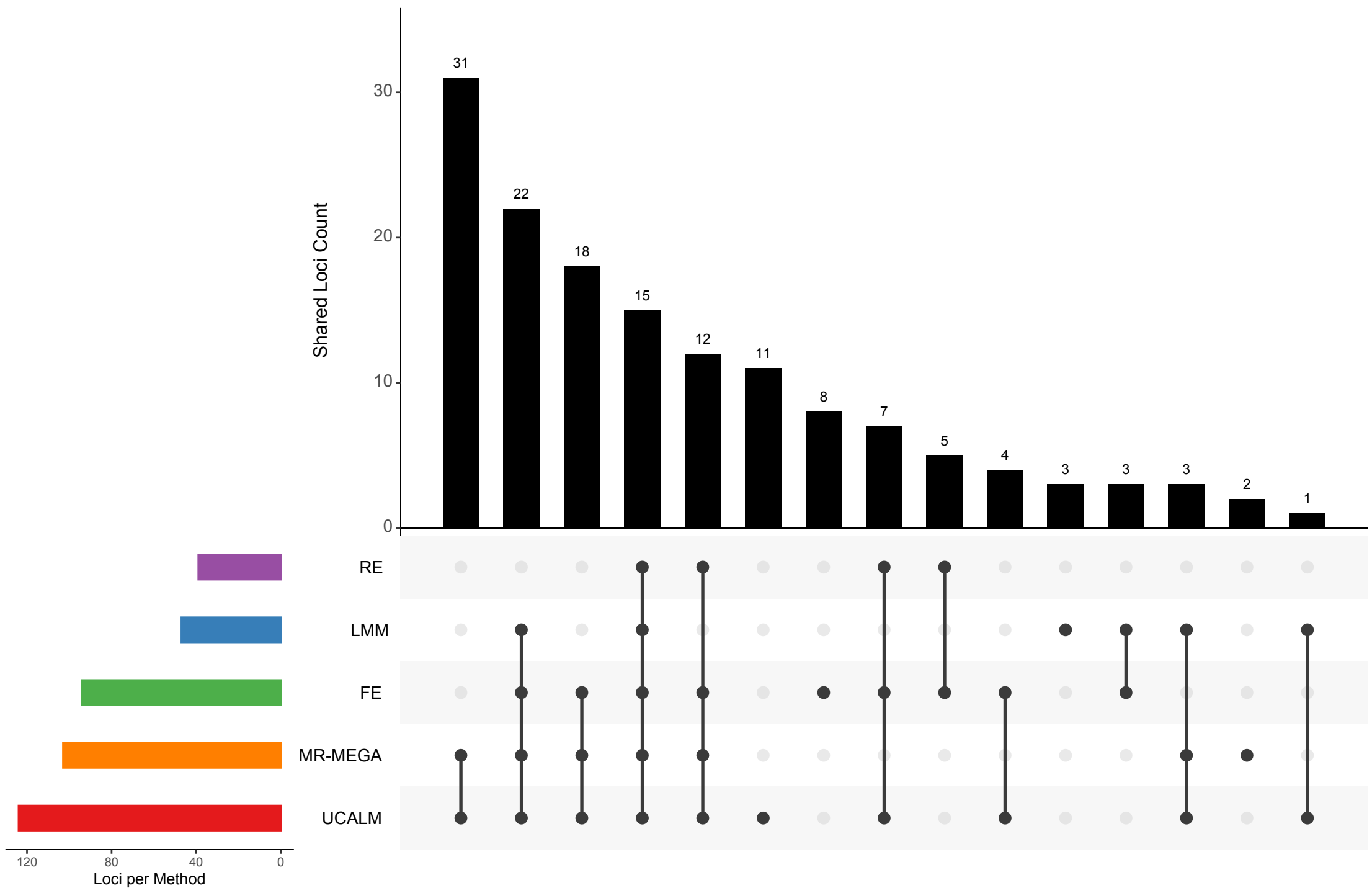

**Supplementary Fig. 11.** UpSet plot of significant loci for the LMA trait across five association strategies. The plot shows the overlap among significant 100 kb genomic windows identified by RE, LMM, MR-MEGA, FE and UCALM. Vertical bars indicate the number of loci in each method intersection, and the connected points below each bar indicate the methods included in that intersection. Horizontal bars show the total number of significant loci detected by each method. Methods are ordered by total locus count, and intersections are ordered by decreasing intersection size.

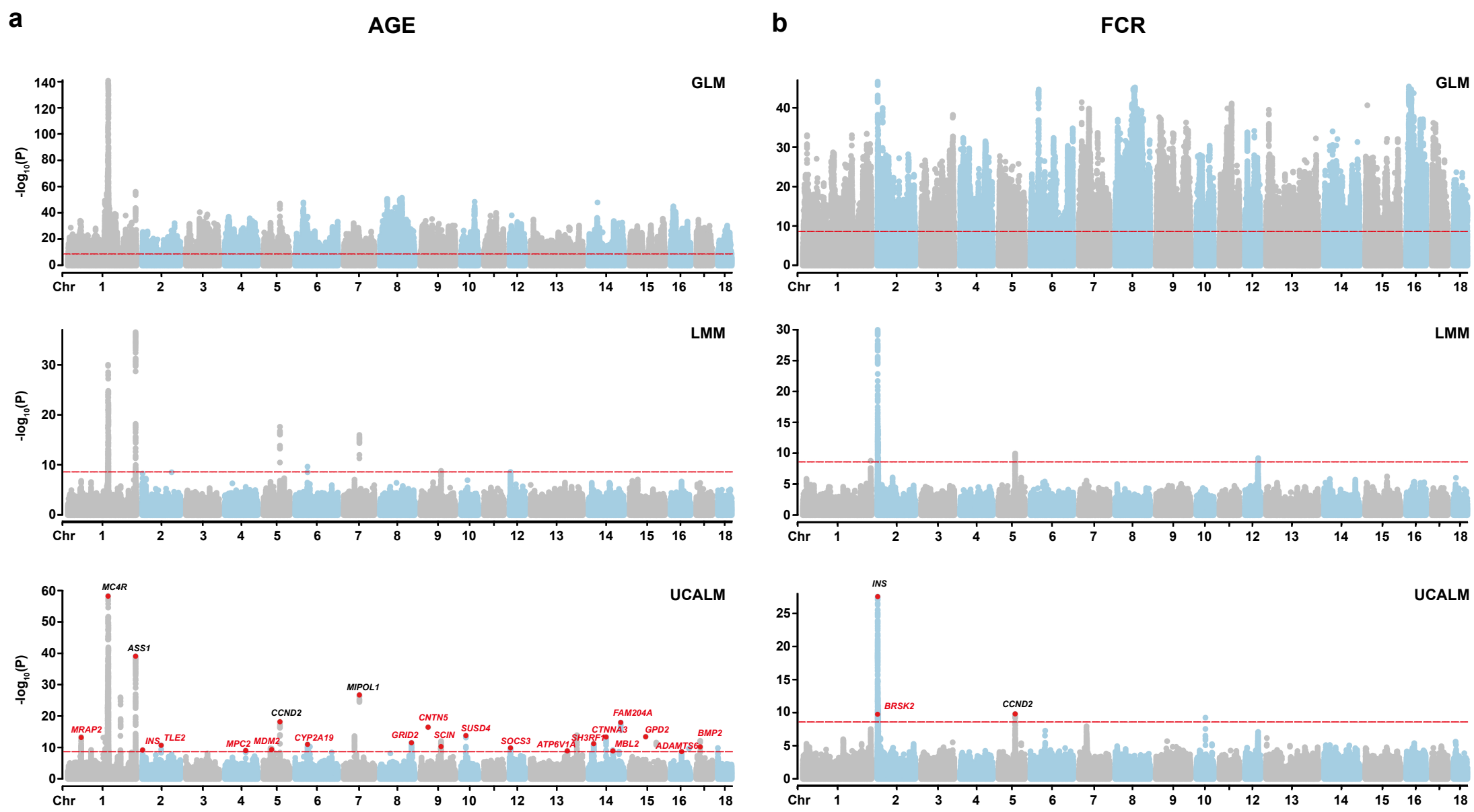

**Supplementary Fig. 12.** Comparison of genome-wide association studies for two pig economic traits across GLM, LMM and UCALM. **a,** Manhattan plots for age adjusted to 100 kg (AGE). **b,** Manhattan plots for feed conversion ratio (FCR). From top to bottom, rows show association results obtained using GLM, LMM and UCALM, respectively. The x-axis indicates chromosomal position, and the y-axis shows association significance as $-\log_{10}(P)$. The red horizontal dashed line indicates the genome-wide significance threshold ($P < 2.54 \times 10^{-9}$). Candidate genes are labelled in the UCALM panels; black labels denote loci detected by both LMM and UCALM, whereas red labels denote loci detected only by UCALM.

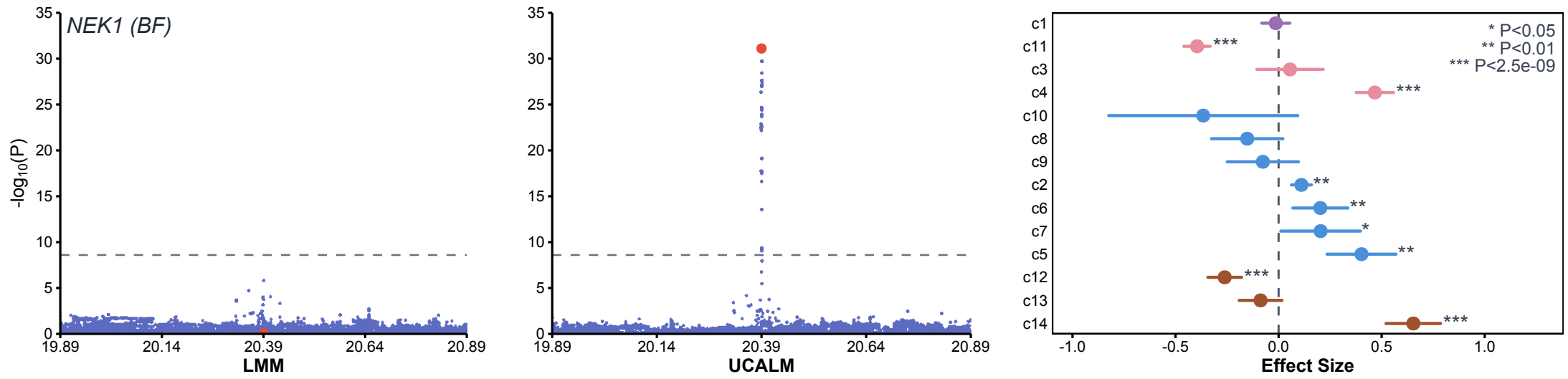

**Supplementary Fig. 13.** Regional association and subgroup effect estimates at the *NEK1* locus for backfat thickness. Regional association plots comparing LMM and UCALM signals at the *NEK1* locus for backfat thickness adjusted to 100 kg. The red point denotes the lead SNP, and the horizontal dashed line indicates the genome-wide significance threshold $(P < 2.54 \times 10^{-9})$. The right panel shows subgroup-level effect estimates for the lead SNP, with points indicating estimated effects and horizontal lines indicating 95% confidence intervals. Point and confidence-interval colors denote the predominant breed composition of each UCALM-derived cluster: Duroc (brown), Landrace (pink), Yorkshire White (blue) and Pietrain (purple). Asterisks indicate nominal or genome-wide significance levels as shown in the figure.

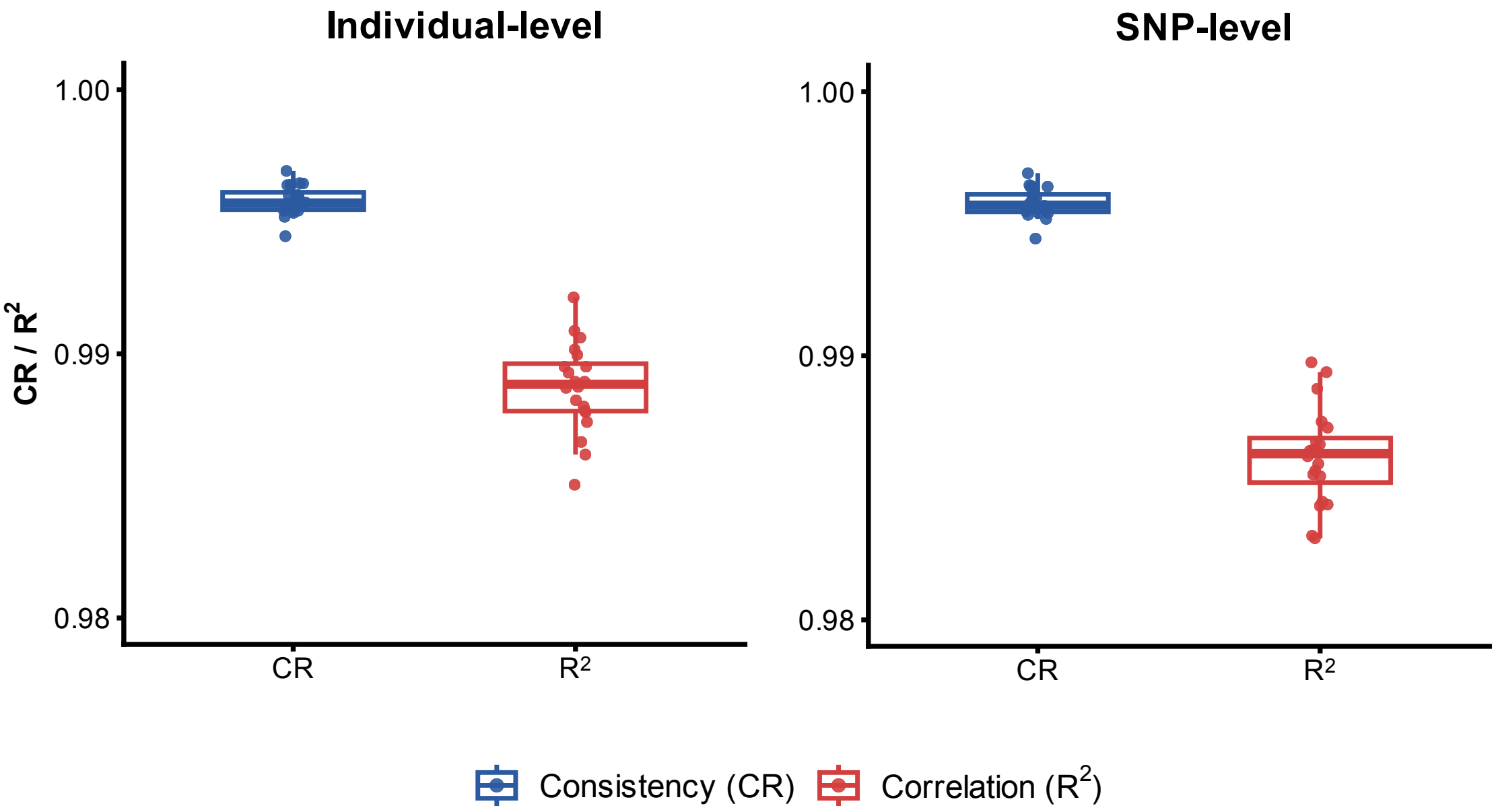

**Supplementary Fig. 14.** Imputation accuracy in the pig dataset assessed by 20-fold cross-validation. **a**, Individual level consistency (CR) and correlation ($R^2$) across the 20 cross-validation folds. **b**, SNP level CR and $R^2$ across the 20 cross-validation folds. Boxplots show the distribution across folds, and points indicate the results of each fold. CR was calculated from hard call genotype concordance, and $R^2$ was calculated from the squared Pearson correlation between imputed and observed genotype dosages.

**Supplementary Table 1.** Mean $\chi^2$ statistics for null SNPs across simulated genetic architectures and heritability levels. Data are presented as the mean $\chi^2$ statistics for null SNPs, with standard deviations (SD) in parentheses, across 20 independent simulation replicates. Columns denote four simulated genetic architectures: HOMO, homogeneous effects; HOMO_HETE, mixed homogeneous and heterogeneous effects; HETE, ancestry-correlated heterogeneous effects; and HETE_onlyEAS, EAS-specific effects. Rows are stratified by trait heritability ($h^2 = 0.2, 0.5, 0.8$). Abbreviations: LMM, linear mixed model; FE, fixed-effect meta-analysis; RE, random-effect meta-analysis; MR-MEGA, meta-regression of multi-ancestry genetic association; UCALM, the proposed method.

| | | **Simulated Genetic Architecture** | | | |
|---|---|---|---|---|---|
| **Heritability** | **Method** | **HOMO** | **HOMO_HETE** | **HETE** | **HETE_onlyEAS** |
| 0.2 | LMM | 0.9070 (0.0069) | 0.9650 (0.0073) | 0.9637 (0.0092) | 1.0033 (0.0103) |
| | FE | 0.9606 (0.0060) | 0.9809 (0.0060) | 0.9973 (0.0074) | 1.0007 (0.0062) |
| | RE | 0.8329 (0.0049) | 0.8398 (0.0049) | 0.8481 (0.0065) | 0.8590 (0.0059) |
| | MR-MEGA | 0.9431 (0.0064) | 0.9705 (0.0065) | 0.9889 (0.0052) | 1.0022 (0.0047) |
| | UCALM | 0.9202 (0.0066) | 0.9496 (0.0060) | 0.9707 (0.0061) | 0.9812 (0.0040) |
| 0.5 | LMM | 0.8217 (0.0060) | 0.9309 (0.0082) | 0.9128 (0.0084) | 1.0003 (0.0048) |
| | FE | 0.9194 (0.0059) | 0.9492 (0.0059) | 0.9815 (0.0063) | 0.9999 (0.0045) |
| | RE | 0.8054 (0.0049) | 0.8157 (0.0046) | 0.8393 (0.0051) | 0.8590 (0.0040) |
| | MR-MEGA | 0.8851 (0.0061) | 0.9236 (0.0078) | 0.9724 (0.0072) | 0.9998 (0.0041) |
| | UCALM | 0.8604 (0.0063) | 0.9025 (0.0071) | 0.9551 (0.0073) | 0.9792 (0.0047) |
| 0.8 | LMM | 0.7445 (0.0037) | 0.9092 (0.0090) | 0.8934 (0.0084) | 1.0026 (0.0190) |
| | FE | 0.8790 (0.0062) | 0.9010 (0.0056) | 0.9687 (0.0048) | 1.0007 (0.0054) |
| | RE | 0.7768 (0.0050) | 0.7790 (0.0045) | 0.9309 (0.0038) | 0.8593 (0.0045) |
| | MR-MEGA | 0.8286 (0.0064) | 0.8635 (0.0081) | 0.9572 (0.0050) | 0.9999 (0.0065) |
| | UCALM | 0.8028 (0.0063) | 0.8386 (0.0071) | 0.9411 (0.0049) | 0.9796 (0.0061) |

**Supplementary Table 2.** False positive rates (FPRs) of different association methods across four genetic architectures at $h^2 = 0.2$ and varying significance levels. Data are presented as pooled FPRs calculated across 20 independent simulation replicates for null variants ($m_{null}$ = 4,410,743). The standard error (SE) of the estimated FPR was computed based on a binomial distribution: $SE = \sqrt{p(1-p)/k}$, where $p$ is the expected nominal FPR (i.e. $0.05, 5 \times 10^{-3}, 5 \times 10^{-4}, and\ 5 \times 10^{-5}$) and $k$ is the total number of tests across all replicates ($k = m_{null} \times 20$). Columns denote four simulated genetic architectures: HOMO, homogeneous effects; HOMO_HETE, mixed homogeneous and heterogeneous effects; HETE, ancestry-correlated heterogeneous effects; and HETE_onlyEAS, EAS-specific effects. Rows are stratified by significance level. Abbreviations: LMM, linear mixed model; FE, fixed-effect meta-analysis; RE, random-effect meta-analysis; MR-MEGA, meta-regression of multi-ancestry genetic association; UCALM, the proposed method.

| Significance Levels | Method | Simulated Genetic Architecture | | | |
|---|---|---|---|---|---|
| | | **HOMO** | **HOMO_HETE** | **HETE** | **HETE_onlyEAS** |
| $P < 0.05$ | LMM | 0.0399 (2.3E-5) | 0.0463 (2.3E-5) | 0.0482 (2.3E-5) | 0.0503 (2.3E-5) |
| | FE | 0.0456 (2.3E-5) | 0.0478 (2.3E-5) | 0.0497 (2.3E-5) | 0.0501 (2.3E-5) |
| | RE | 0.0339 (2.3E-5) | 0.0350 (2.3E-5) | 0.0363 (2.3E-5) | 0.0369 (2.3E-5) |
| | MR-MEGA | 0.0444 (2.3E-5) | 0.0471 (2.3E-5) | 0.0489 (2.3E-5) | 0.0502 (2.3E-5) |
| | UCALM | 0.0471 (2.3E-5) | 0.0501 (2.3E-5) | 0.0522 (2.3E-5) | 0.0533 (2.3E-5) |
| $P < 5 \times 10^{-3}$ | LMM | 3.3E-03 (7.5E-6) | 4.8E-03 (7.5E-6) | 8.6E-03 (7.5E-6) | 5.1E-03 (7.5E-6) |
| | FE | 4.2E-03 (7.5E-6) | 4.6E-03 (7.5E-6) | 4.9E-03 (7.5E-6) | 5.0E-03 (7.5E-6) |
| | RE | 2.9E-03 (7.5E-6) | 3.2E-03 (7.5E-6) | 3.4E-03 (7.5E-6) | 3.4E-03 (7.5E-6) |
| | MR-MEGA | 4.1E-03 (7.5E-6) | 4.6E-03 (7.5E-6) | 4.7E-03 (7.5E-6) | 5.0E-03 (7.5E-6) |
| | UCALM | 4.3E-03 (7.5E-6) | 4.8E-03 (7.5E-6) | 5.1E-03 (7.5E-6) | 5.3E-03 (7.5E-6) |
| $P < 5 \times 10^{-4}$ | LMM | 2.8E-04 (2.4E-6) | 5.6E-04 (2.4E-6) | 2.5E-04 (2.4E-6) | 5.1E-04 (2.4E-6) |
| | FE | 3.7E-04 (2.4E-6) | 4.4E-04 (2.4E-6) | 5.0E-04 (2.4E-6) | 5.0E-04 (2.4E-6) |
| | RE | 2.5E-04 (2.4E-6) | 2.9E-04 (2.4E-6) | 3.3E-04 (2.4E-6) | 3.3E-04 (2.4E-6) |
| | MR-MEGA | 3.9E-04 (2.4E-6) | 4.4E-04 (2.4E-6) | 4.6E-04 (2.4E-6) | 5.0E-04 (2.4E-6) |
| | UCALM | 3.8E-04 (2.4E-6) | 4.6E-04 (2.4E-6) | 4.9E-04 (2.4E-6) | 5.1E-04 (2.4E-6) |
| $P < 5 \times 10^{-5}$ | LMM | 2.6E-05 (7.5E-7) | 8.9E-05 (7.5E-7) | 9.9E-04 (7.5E-7) | 4.8E-05 (7.5E-7) |
| | FE | 3.3E-05 (7.5E-7) | 4.3E-05 (7.5E-7) | 5.2E-05 (7.5E-7) | 4.7E-05 (7.5E-7) |
| | RE | 2.4E-05 (7.5E-7) | 2.9E-05 (7.5E-7) | 3.3E-05 (7.5E-7) | 2.8E-05 (7.5E-7) |
| | MR-MEGA | 3.8E-05 (7.5E-7) | 4.2E-05 (7.5E-7) | 4.5E-05 (7.5E-7) | 4.8E-05 (7.5E-7) |
| | UCALM | 3.6E-05 (7.5E-7) | 4.4E-05 (7.5E-7) | 5.1E-05 (7.5E-7) | 4.7E-05 (7.5E-7) |

**Supplementary Table 3.** False positive rates (FPRs) of different association methods across four genetic architectures at $h^2 = 0.5$ and varying significance levels. Data are presented as pooled FPRs calculated across 20 independent simulation replicates for null variants ($m_{null} = 4{,}410{,}743$). The standard error (SE) of the estimated FPR was computed based on a binomial distribution: $SE = \sqrt{p(1-p)/k}$, where $p$ is the expected nominal FPR (i.e. $0.05, 5 \times 10^{-3}, 5 \times 10^{-4}, and\ 5 \times 10^{-5}$) and $k$ is the total number of tests across all replicates ($k = m_{null} \times 20$). Columns denote four simulated genetic architectures: HOMO, homogeneous effects; HOMO_HETE, mixed homogeneous and heterogeneous effects; HETE, ancestry-correlated heterogeneous effects; and HETE_onlyEAS, EAS-specific effects. Rows are stratified by significance level. Abbreviations: LMM, linear mixed model; FE, fixed-effect meta-analysis; RE, random-effect meta-analysis; MR-MEGA, meta-regression of multi-ancestry genetic association; UCALM, the proposed method.

| **Significance Levels** | **Method** | **Simulated Genetic Architecture** | | | |
|---|---|---|---|---|---|
| | | **HOMO** | **HOMO_HETE** | **HETE** | **HETE_onlyEAS** |
| | LMM | 0.0316 (2.3E-5) | 0.0433 (2.3E-5) | 0.0447 (2.3E-5) | 0.0501 (2.3E-5) |
| | FE | 0.0412 (2.3E-5) | 0.0443 (2.3E-5) | 0.0480 (2.3E-5) | 0.0501 (2.3E-5) |
| $P < 0.05$ | RE | 0.0311 (2.3E-5) | 0.0326 (2.3E-5) | 0.0353 (2.3E-5) | 0.0369 (2.3E-5) |
| | MR-MEGA | 0.0392 (2.3E-5) | 0.0427 (2.3E-5) | 0.0472 (2.3E-5) | 0.0500 (2.3E-5) |
| | UCALM | 0.0413 (2.3E-5) | 0.0454 (2.3E-5) | 0.0505 (2.3E-5) | 0.0532 (2.3E-5) |
| | LMM | 2.3E-03 (7.5E-6) | 5.9E-03 (7.5E-6) | 5.9E-03 (7.5E-6) | 5.0E-03 (7.5E-6) |
| | FE | 3.6E-03 (7.5E-6) | 4.1E-03 (7.5E-6) | 4.6E-03 (7.5E-6) | 5.0E-03 (7.5E-6) |
| $P < 5 \times 10^{-3}$ | RE | 2.5E-03 (7.5E-6) | 2.8E-03 (7.5E-6) | 3.2E-03 (7.5E-6) | 3.4E-03 (7.5E-6) |
| | MR-MEGA | 3.5E-03 (7.5E-6) | 3.9E-03 (7.5E-6) | 4.6E-03 (7.5E-6) | 5.0E-03 (7.5E-6) |
| | UCALM | 3.6E-03 (7.5E-6) | 4.1E-03 (7.5E-6) | 4.9E-03 (7.5E-6) | 5.3E-03 (7.5E-6) |
| | LMM | 1.8E-04 (2.4E-6) | 1.3E-03 (2.4E-6) | 1.1E-03 (2.4E-6) | 4.9E-04 (2.4E-6) |
| | FE | 3.1E-04 (2.4E-6) | 3.7E-04 (2.4E-6) | 4.6E-04 (2.4E-6) | 4.9E-04 (2.4E-6) |
| $P < 5 \times 10^{-4}$ | RE | 2.2E-04 (2.4E-6) | 2.5E-04 (2.4E-6) | 3.1E-04 (2.4E-6) | 3.3E-04 (2.4E-6) |
| | MR-MEGA | 3.1E-04 (2.4E-6) | 3.7E-04 (2.4E-6) | 4.5E-04 (2.4E-6) | 5.0E-04 (2.4E-6) |
| | UCALM | 3.1E-04 (2.4E-6) | 3.8E-04 (2.4E-6) | 4.7E-04 (2.4E-6) | 5.1E-04 (2.4E-6) |
| | LMM | 1.5E-05 (7.5E-7) | 4.2E-04 (7.5E-7) | 3.0E-04 (7.5E-7) | 4.7E-05 (7.5E-7) |
| | FE | 2.9E-05 (7.5E-7) | 3.5E-05 (7.5E-7) | 4.5E-05 (7.5E-7) | 4.7E-05 (7.5E-7) |
| $P < 5 \times 10^{-5}$ | RE | 2.0E-05 (7.5E-7) | 2.3E-05 (7.5E-7) | 2.9E-05 (7.5E-7) | 3.1E-05 (7.5E-7) |
| | MR-MEGA | 2.7E-05 (7.5E-7) | 3.5E-05 (7.5E-7) | 4.3E-05 (7.5E-7) | 4.9E-05 (7.5E-7) |
| | UCALM | 2.6E-05 (7.5E-7) | 3.6E-05 (7.5E-7) | 4.6E-05 (7.5E-7) | 4.8E-05 (7.5E-7) |

**Supplementary Table 4.** False positive rates (FPRs) of different association methods across four genetic architectures at $h^2 = 0.8$ and varying significance levels. Data are presented as pooled FPRs calculated across 20 independent simulation replicates for null variants ($m_{null}$ = 4,410,743). The standard error (SE) of the estimated FPR was computed based on a binomial distribution: $SE = \sqrt{p(1-p)/k}$, where $p$ is the expected nominal FPR (i.e. $0.05, 5 \times 10^{-3}, 5 \times 10^{-4}, and\ 5 \times 10^{-5}$) and $k$ is the total number of tests across all replicates ($k = m_{null} \times 20$). Columns denote four simulated genetic architectures: HOMO, homogeneous effects; HOMO_HETE, mixed homogeneous and heterogeneous effects; HETE, ancestry-correlated heterogeneous effects; and HETE_onlyEAS, EAS-specific effects. Rows are stratified by significance level. Abbreviations: LMM, linear mixed model; FE, fixed-effect meta-analysis; RE, random-effect meta-analysis; MR-MEGA, meta-regression of multi-ancestry genetic association; UCALM, the proposed method.

| Significance Levels | Method | Simulated Genetic Architecture: HOMO | HOMO_HETE | HETE | HETE_onlyEAS |
|---|---|---|---|---|---|
| | LMM | 0.0248 (2.3E-5) | 0.0424 (2.3E-5) | 0.0438 (2.3E-5) | 0.0502 (2.3E-5) |
| | FE | 0.0371 (2.3E-5) | 0.0393 (2.3E-5) | 0.0465 (2.3E-5) | 0.0500 (2.3E-5) |
| $P < 0.05$ | RE | 0.0282 (2.3E-5) | 0.0289 (2.3E-5) | 0.0341 (2.3E-5) | 0.0368 (2.3E-5) |
| | MR-MEGA | 0.0347 (2.3E-5) | 0.0375 (2.3E-5) | 0.0457 (2.3E-5) | 0.0500 (2.3E-5) |
| | UCALM | 0.0364 (2.3E-5) | 0.0394 (2.3E-5) | 0.0489 (2.3E-5) | 0.0532 (2.3E-5) |
| | LMM | 1.6E-03 (7.5E-6) | 7.9E-03 (7.5E-6) | 5.1E-03 (7.5E-6) | 5.1E-03 (7.5E-6) |
| | FE | 3.0E-03 (7.5E-6) | 3.3E-03 (7.5E-6) | 4.4E-03 (7.5E-6) | 5.0E-03 (7.5E-6) |
| $P < 5 \times 10^{-3}$ | RE | 2.1E-03 (7.5E-6) | 2.3E-03 (7.5E-6) | 3.0E-03 (7.5E-6) | 3.5E-03 (7.5E-6) |
| | MR-MEGA | 2.9E-03 (7.5E-6) | 3.2E-03 (7.5E-6) | 4.3E-03 (7.5E-6) | 5.0E-03 (7.5E-6) |
| | UCALM | 3.0E-03 (7.5E-6) | 3.4E-03 (7.5E-6) | 4.6E-03 (7.5E-6) | 5.4E-03 (7.5E-6) |
| | LMM | 1.1E-04 (2.4E-6) | 2.5E-03 (2.4E-6) | 7.4E-04 (2.4E-6) | 5.1E-04 (2.4E-6) |
| | FE | 2.4E-04 (2.4E-6) | 2.9E-04 (2.4E-6) | 4.0E-04 (2.4E-6) | 5.0E-04 (2.4E-6) |
| $P < 5 \times 10^{-4}$ | RE | 1.7E-04 (2.4E-6) | 2.0E-04 (2.4E-6) | 2.7E-04 (2.4E-6) | 3.3E-04 (2.4E-6) |
| | MR-MEGA | 2.4E-04 (2.4E-6) | 2.9E-04 (2.4E-6) | 4.1E-04 (2.4E-6) | 5.0E-04 (2.4E-6) |
| | UCALM | 2.4E-04 (2.4E-6) | 2.9E-04 (2.4E-6) | 4.2E-04 (2.4E-6) | 5.2E-04 (2.4E-6) |
| | LMM | 7.5E-06 (7.5E-7) | 1.1E-03 (7.5E-7) | 1.6E-04 (7.5E-7) | 4.7E-05 (7.5E-7) |
| | FE | 2.0E-05 (7.5E-7) | 2.6E-05 (7.5E-7) | 3.8E-05 (7.5E-7) | 4.7E-05 (7.5E-7) |
| $P < 5 \times 10^{-5}$ | RE | 1.5E-05 (7.5E-7) | 1.7E-05 (7.5E-7) | 2.7E-05 (7.5E-7) | 3.0E-05 (7.5E-7) |
| | MR-MEGA | 2.0E-05 (7.5E-7) | 2.7E-05 (7.5E-7) | 3.8E-05 (7.5E-7) | 4.8E-05 (7.5E-7) |
| | UCALM | 2.0E-05 (7.5E-7) | 2.6E-05 (7.5E-7) | 3.9E-05 (7.5E-7) | 4.9E-05 (7.5E-7) |

**Supplementary Table 5.** Mean $\chi^2$ statistics for causal SNPs across simulated genetic architectures and heritability levels. Data are presented as mean $\chi^2$ statistics for causal SNPs, with standard deviations (SDs) in parentheses, across 20 independent simulation replicates. Columns denote four simulated genetic architectures: HOMO, homogeneous effects; HOMO_HETE, mixed homogeneous and heterogeneous effects; HETE, ancestry-correlated heterogeneous effects; and HETE_onlyEAS, EAS-specific effects. Rows are stratified by trait heritability ($h^2 = 0.2,0.5,0.8$). Abbreviations: LMM, linear mixed model; FE, fixed-effect meta-analysis; RE, random-effect meta-analysis; MR-MEGA, meta-regression of multi-ancestry genetic association; UCALM, the proposed method.

| | | **Simulated Genetic Architecture** | | | |
|---|---|---|---|---|---|
| **Heritability** | **Method** | **HOMO** | **HOMO_HETE** | **HETE** | **HETE_onlyEAS** |
| 0.2 | LMM | 9.64 (0.28) | 5.96 (0.55) | 5.28 (1.26) | 1.32 (0.17) |
| | FE | 10.55 (0.37) | 4.40 (0.18) | 1.19 (0.13) | 1.30 (0.16) |
| | RE | 8.77 (0.34) | 3.44 (0.14) | 0.63 (0.08) | 1.00 (0.14) |
| | MR-MEGA | 8.31 (0.33) | 6.55 (1.43) | 12.89 (3.47) | 1.55 (0.20) |
| | UCALM | 9.78 (0.36) | 7.44 (1.42) | 12.21 (3.45) | 1.83 (0.24) |
| 0.5 | LMM | 21.19 (0.62) | 12.75 (1.15) | 6.35 (1.80) | 1.73 (0.23) |
| | FE | 24.14 (0.60) | 10.61 (0.39) | 1.73 (0.24) | 1.70 (0.24) |
| | RE | 20.11 (0.51) | 8.39 (0.29) | 0.71 (0.08) | 1.09 (0.14) |
| | MR-MEGA | 20.50 (0.59) | 19.34 (4.00) | 29.72 (7.67) | 2.69 (0.50) |
| | UCALM | 23.04 (0.60) | 20.18 (3.99) | 28.80 (7.64) | 4.30 (0.51) |
| 0.8 | LMM | 35.02 (0.54) | 19.10 (1.63) | 6.69 (1.70) | 2.11 (0.40) |
| | FE | 39.29 (0.91) | 21.49 (0.67) | 2.26 (0.38) | 2.05 (0.39) |
| | RE | 32.44 (0.93) | 17.04 (0.58) | 0.79 (0.10) | 1.09 (0.14) |
| | MR-MEGA | 34.72 (0.92) | 37.71 (6.61) | 46.87 (12.31) | 3.78 (0.47) |
| | UCALM | 38.03 (0.92) | 38.95 (6.62) | 45.82 (12.31) | 7.26 (0.49) |

**Supplementary Table 6.** Mean number of detected causal SNPs across simulated genetic architectures and heritability levels. Data are presented as the mean number of causal SNPs reaching the Bonferroni significance threshold, with standard deviations (SDs) in parentheses, across 20 independent simulation replicates. Columns denote four simulated genetic architectures: HOMO, homogeneous effects; HOMO_HETE, mixed homogeneous and heterogeneous effects; HETE, ancestry-correlated heterogeneous effects; and HETE_onlyEAS, EAS-specific effects. Rows are stratified by trait heritability ($h^2 = 0.2, 0.5, 0.8$). Abbreviations: LMM, linear mixed model; FE, fixed-effect meta-analysis; RE, random-effect meta-analysis; MR-MEGA, meta-regression of multi-ancestry genetic association; UCALM, the proposed method.

| | | **Simulated Genetic Architecture** | | | |
|---|---|---|---|---|---|
| **Heritability** | **Method** | **HOMO** | **HOMO_HETE** | **HETE** | **HETE_onlyEAS** |
| 0.2 | LMM | 68.65 (5.88) | 28.55 (8.30) | 21.35 (19.67) | 0.00 (0.00) |
| | FE | 77.35 (6.85) | 20.85 (3.03) | 0.00 (0.00) | 0.00 (0.00) |
| | RE | 59.85 (4.49) | 14.90 (2.94) | 0.00 (0.00) | 0.00 (0.00) |
| | MR-MEGA | 58.25 (5.62) | 34.95 (10.87) | 97.35 (30.26) | 0.00 (0.00) |
| | UCALM | 72.25 (6.91) | 44.80 (11.23) | 91.80 (28.38) | 0.05 (0.22) |
| 0.5 | LMM | 180.80 (10.46) | 104.45 (15.96) | 31.90 (21.93) | 0.00 (0.00) |
| | FE | 201.80 (11.47) | 84.15 (8.67) | 1.75 (3.39) | 0.00 (0.00) |
| | RE | 166.25 (9.07) | 65.35 (5.50) | 0.00 (0.00) | 0.00 (0.00) |
| | MR-MEGA | 169.25 (9.22) | 151.40 (21.72) | 217.25 (27.86) | 0.25 (0.55) |
| | UCALM | 191.25 (11.11) | 159.85 (21.63) | 210.15 (26.83) | 1.60 (1.14) |
| 0.8 | LMM | 281.80 (13.56) | 164.25 (19.28) | 35.80 (21.41) | 0.05 (0.22) |
| | FE | 299.15 (12.43) | 155.65 (7.73) | 2.80 (3.58) | 0.00 (0.00) |
| | RE | 251.00 (15.01) | 126.95 (7.44) | 0.00 (0.00) | 0.00 (0.00) |
| | MR-MEGA | 258.80 (13.29) | 264.40 (17.45) | 287.45 (40.67) | 0.95 (0.94) |
| | UCALM | 287.55 (11.23) | 274.85 (18.28) | 279.70 (39.80) | 5.40 (1.67) |

**Supplementary Table 7.** Genomic inflation factors ($\lambda_{GC}$) stratified by minor allele frequency across 24 human complex traits. Genomic inflation factors were calculated to benchmark population stratification control across GLM, UCALM, and LDAK-KVIK. To evaluate model robustness, $\lambda_{GC}$ values were computed for rare ($0.001 \leq \text{MAF} < 0.01$), low-frequency ($0.01 \leq \text{MAF} < 0.05$), and common ($0.05 \leq \text{MAF} \leq 0.5$) variants, and the full dataset (All). The first column lists the trait abbreviations, and the second column indicates the MAF category. The bottom rows summarize the mean $\lambda_{GC}$ across all 24 traits for each MAF category and method.

| Traits Abbr. | MAF Category | GLM | UCALM | LDAK.KVIK |
|---|---|---|---|---|
| AFLB | Rare [0.001, 0.01) | 1.1421 | 0.8869 | 1.0192 |
| | Low-frequency [0.01, 0.05) | 1.1713 | 0.9658 | 1.1084 |
| | Common [0.05, 0.5] | 1.5262 | 1.2759 | 1.4470 |
| | All | 1.2849 | 1.0366 | 1.1851 |
| AMena | Rare [0.001, 0.01) | 1.2139 | 0.8984 | 1.0961 |
| | Low-frequency [0.01, 0.05) | 1.2614 | 1.0377 | 1.2314 |
| | Common [0.05, 0.5] | 1.7317 | 1.4779 | 1.7588 |
| | All | 1.4020 | 1.1147 | 1.3411 |
| AMeno | Rare [0.001, 0.01) | 1.0203 | 0.8573 | 0.9481 |
| | Low-frequency [0.01, 0.05) | 1.0783 | 0.9040 | 1.0533 |
| | Common [0.05, 0.5] | 1.2402 | 1.0297 | 1.2277 |
| | All | 1.1118 | 0.9296 | 1.0682 |
| BFP | Rare [0.001, 0.01) | 1.1206 | 0.9460 | 0.9936 |
| | Low-frequency [0.01, 0.05) | 1.5038 | 1.2645 | 1.4400 |
| | Common [0.05, 0.5] | 2.5058 | 2.1686 | 2.5437 |
| | All | 1.5968 | 1.3615 | 1.5015 |
| BMI | Rare [0.001, 0.01) | 1.2254 | 0.9620 | 1.1133 |
| | Low-frequency [0.01, 0.05) | 1.5299 | 1.3076 | 1.4862 |
| | Common [0.05, 0.5] | 2.5778 | 2.2625 | 2.6624 |
| | All | 1.6873 | 1.4036 | 1.6190 |
| BMR | Rare [0.001, 0.01) | 1.2262 | 0.9765 | 1.1044 |
| | Low-frequency [0.01, 0.05) | 1.5380 | 1.3178 | 1.5353 |
| | Common [0.05, 0.5] | 2.5895 | 2.2804 | 2.7827 |
| | All | 1.6875 | 1.4170 | 1.6465 |
| BW | Rare [0.001, 0.01) | 1.0639 | 0.8752 | 0.9865 |
| | Low-frequency [0.01, 0.05) | 1.1476 | 0.9449 | 1.1213 |
| | Common [0.05, 0.5] | 1.3825 | 1.1468 | 1.3628 |
| | All | 1.1982 | 0.9842 | 1.1454 |
| DBP | Rare [0.001, 0.01) | 1.1338 | 0.9159 | 1.0452 |
| | Low-frequency [0.01, 0.05) | 1.2603 | 1.0401 | 1.2209 |
| | Common [0.05, 0.5] | 1.8024 | 1.5278 | 1.7966 |
| | All | 1.3800 | 1.1387 | 1.3211 |
| EA | Rare [0.001, 0.01) | 1.6286 | 0.8865 | 1.4801 |
| | Low-frequency [0.01, 0.05) | 1.1667 | 0.9250 | 1.1178 |
| | Common [0.05, 0.5] | 1.3616 | 1.1538 | 1.3282 |
| | All | 1.4290 | 0.9896 | 1.3496 |

| Traits Abbr. | MAF Category | GLM | UCALM | LDAK.KVIK |
|---|---|---|---|---|
| FEV | Rare [0.001, 0.01) | 1.1120 | 0.9347 | 0.9933 |
| | Low-frequency [0.01, 0.05) | 1.3463 | 1.1241 | 1.2918 |
| | Common [0.05, 0.5] | 2.1089 | 1.8076 | 2.1157 |
| | All | 1.4645 | 1.2427 | 1.3843 |
| FIS | Rare [0.001, 0.01) | 0.9128 | 0.8950 | 0.8086 |
| | Low-frequency [0.01, 0.05) | 1.1715 | 1.0113 | 1.0933 |
| | Common [0.05, 0.5] | 1.6048 | 1.3600 | 1.5435 |
| | All | 1.1883 | 1.0773 | 1.0962 |
| FVC | Rare [0.001, 0.01) | 1.0924 | 0.8747 | 0.9708 |
| | Low-frequency [0.01, 0.05) | 1.3221 | 1.1408 | 1.2688 |
| | Common [0.05, 0.5] | 2.0176 | 1.7347 | 2.0084 |
| | All | 1.4232 | 1.1919 | 1.3393 |
| hBMD | Rare [0.001, 0.01) | 1.4674 | 0.9148 | 1.4159 |
| | Low-frequency [0.01, 0.05) | 1.2558 | 1.0471 | 1.2463 |
| | Common [0.05, 0.5] | 1.5702 | 1.3382 | 1.6370 |
| | All | 1.4930 | 1.0838 | 1.4944 |
| HC | Rare [0.001, 0.01) | 1.1883 | 0.9433 | 1.0919 |
| | Low-frequency [0.01, 0.05) | 1.4450 | 1.2384 | 1.4085 |
| | Common [0.05, 0.5] | 2.3630 | 2.0452 | 2.4440 |
| | All | 1.5937 | 1.3279 | 1.5431 |
| HGSL | Rare [0.001, 0.01) | 1.1849 | 0.9097 | 1.0709 |
| | Low-frequency [0.01, 0.05) | 1.2906 | 1.0728 | 1.2338 |
| | Common [0.05, 0.5] | 1.8515 | 1.5837 | 1.8308 |
| | All | 1.4286 | 1.1591 | 1.3480 |
| HGSR | Rare [0.001, 0.01) | 1.1992 | 0.9178 | 1.0854 |
| | Low-frequency [0.01, 0.05) | 1.2891 | 1.0589 | 1.2381 |
| | Common [0.05, 0.5] | 1.8422 | 1.5728 | 1.8226 |
| | All | 1.4329 | 1.1553 | 1.3546 |
| HT | Rare [0.001, 0.01) | 1.2267 | 1.0471 | 1.1565 |
| | Low-frequency [0.01, 0.05) | 1.7567 | 1.5142 | 1.8328 |
| | Common [0.05, 0.5] | 2.9618 | 2.6690 | 3.4124 |
| | All | 1.8096 | 1.5837 | 1.8569 |
| MTCIM | Rare [0.001, 0.01) | 1.4213 | 0.8865 | 1.3112 |
| | Low-frequency [0.01, 0.05) | 1.1830 | 0.9451 | 1.1396 |
| | Common [0.05, 0.5] | 1.4953 | 1.2642 | 1.4681 |
| | All | 1.4010 | 1.0293 | 1.3355 |
| NTS | Rare [0.001, 0.01) | 1.0335 | 0.8828 | 0.9583 |
| | Low-frequency [0.01, 0.05) | 1.2179 | 1.0026 | 1.1883 |
| | Common [0.05, 0.5] | 1.7086 | 1.4038 | 1.6853 |
| | All | 1.2901 | 1.0771 | 1.2342 |

| Traits Abbr. | MAF Category | GLM | UCALM | LDAK.KVIK |
|---|---|---|---|---|
| PEF | Rare [0.001, 0.01) | 1.1326 | 0.8909 | 1.0303 |
| | Low-frequency [0.01, 0.05) | 1.1957 | 0.9860 | 1.1479 |
| | Common [0.05, 0.5] | 1.5945 | 1.3452 | 1.5771 |
| | All | 1.3046 | 1.0625 | 1.2364 |
| PR | Rare [0.001, 0.01) | 1.0413 | 0.8688 | 0.9691 |
| | Low-frequency [0.01, 0.05) | 1.0943 | 0.9120 | 1.0356 |
| | Common [0.05, 0.5] | 1.2678 | 1.0586 | 1.2420 |
| | All | 1.1354 | 0.9479 | 1.0815 |
| SBP | Rare [0.001, 0.01) | 1.0889 | 0.8995 | 0.9927 |
| | Low-frequency [0.01, 0.05) | 1.2689 | 1.0461 | 1.2288 |
| | Common [0.05, 0.5] | 1.8603 | 1.5703 | 1.8586 |
| | All | 1.3735 | 1.1432 | 1.3090 |
| WC | Rare [0.001, 0.01) | 1.1723 | 0.9418 | 1.0649 |
| | Low-frequency [0.01, 0.05) | 1.4390 | 1.2234 | 1.3917 |
| | Common [0.05, 0.5] | 2.3453 | 2.0269 | 2.3785 |
| | All | 1.5786 | 1.3181 | 1.5056 |
| WT | Rare [0.001, 0.01) | 1.2246 | 0.9689 | 1.0926 |
| | Low-frequency [0.01, 0.05) | 1.5213 | 1.3101 | 1.5033 |
| | Common [0.05, 0.5] | 2.5993 | 2.2774 | 2.7403 |
| | All | 1.6860 | 1.4126 | 1.6227 |
| **Mean (Rare)** | \ | **1.1780** | **0.9159** | **1.0750** |
| **Mean (Low-freq)** | \ | **1.3106** | **1.0975** | **1.2735** |
| **Mean (Common)** | \ | **1.9129** | **1.6409** | **1.9448** |
| **Mean (All)** | \ | **1.4325** | **1.1745** | **1.3716** |

**Supplementary Table 8.** Genomic inflation factors ($\lambda_{GC}$) stratified by minor allele frequency across 4 pig traits. Genomic inflation factors were calculated to benchmark population stratification control across GLM, LMM and UCALM. To evaluate model robustness, MAF was recalculated among individuals with available phenotypic records for each trait, and $\lambda_{GC}$ values were computed for rare ($0.001 \leq \text{MAF} < 0.01$), low-frequency ($0.01 \leq \text{MAF} < 0.05$), and common ($0.05 \leq \text{MAF} \leq 0.5$) variants, and the full dataset (All). The first column lists the trait abbreviations, and the second column indicates the MAF category. The bottom rows summarize the mean $\lambda_{GC}$ across all 4 traits for each MAF category and method.

| Traits Abbr. | MAF Category | GLM | LMM | UCALM |
|---|---|---|---|---|
| BF | Rare [0.001, 0.01) | 7.8376 | 0.8925 | 0.8252 |
| | Low-frequency [0.01, 0.05) | 22.7962 | 0.9517 | 0.9036 |
| | Common [0.05, 0.5] | 29.5414 | 0.8640 | 0.9479 |
| | All | 27.6157 | 0.8835 | 0.9358 |
| AGE | Rare [0.001, 0.01) | 5.7955 | 1.0337 | 0.7495 |
| | Low-frequency [0.01, 0.05) | 7.2117 | 1.0268 | 0.9265 |
| | Common [0.05, 0.5] | 8.8343 | 0.9222 | 0.9108 |
| | All | 8.3870 | 0.9460 | 0.9138 |
| FCR | Rare [0.001, 0.01) | 7.1875 | 1.1107 | 1.0261 |
| | Low-frequency [0.01, 0.05) | 5.9279 | 1.0063 | 0.9012 |
| | Common [0.05, 0.5] | 7.3382 | 0.9522 | 0.8624 |
| | All | 6.9906 | 0.9652 | 0.8713 |
| LMA | Rare [0.001, 0.01) | 7.0147 | 1.1896 | 1.1647 |
| | Low-frequency [0.01, 0.05) | 7.9346 | 1.0061 | 0.9731 |
| | Common [0.05, 0.5] | 13.7743 | 0.9057 | 0.9799 |
| | All | 11.8137 | 0.9376 | 0.9854 |
| **Mean (Rare)** | \ | **6.9589** | **1.0566** | **0.9414** |
| **Mean (Low-freq)** | \ | **10.9677** | **0.9977** | **0.9261** |
| **Mean (Common)** | \ | **14.8721** | **0.9110** | **0.9253** |
| **Mean (All)** | \ | **13.7018** | **0.9331** | **0.9266** |

**Supplementary Table 9.** Abbreviated names and details of the 24 quantitative traits in the UKB dataset. The first column is the abbreviation, the second column is the UKB data-field ID of each trait, the third column is the number of records available for analysis, the fourth column indicates the category of the phenotype, and the last column contains the full names of these traits.

| Traits Abbr. | Data-field | Count | Category | Description |
|---|---|---|---|---|
| AFLB | 2754 | 178,615 | Female-specific factors | Age at first live birth |
| AMena | 2714 | 255,657 | Female-specific factors | Age when periods started (menarche) |
| AMeno | 3581 | 149,511 | Female-specific factors | Age at menopause (last menstrual period) |
| BFP | 23099 | 477,932 | Body composition by impedance | Body fat percentage |
| BMI | 21001 | 484,961 | Body size measures | Body mass index |
| BMR | 23105 | 478,185 | Body composition by impedance | Basal metabolic rate |
| BW | 20022 | 268,845 | Early life factors | Birth weight |
| DBP | 4079 | 455,800 | Blood pressure | Diastolic blood pressure, automated reading |
| EA | 845 | 320,389 | Education | Age completed full time education |
| FEV | 3063 | 443,673 | Spirometry | Forced expiratory volume in 1-second |
| FIS | 20016 | 159,989 | Fluid intelligence / reasoning | Fluid intelligence score |
| FVC | 3062 | 443,673 | Spirometry | Forced vital capacity |
| hBMD | 78 | 274,253 | Bone-densitometry of heel | Heel bone mineral density (BMD) T-score, automated |
| HC | 49 | 485,806 | Body size measures | Hip circumference |
| HGSL | 46 | 484,654 | Hand grip strength | Hand grip strength(left) |
| HGSR | 47 | 484,725 | Hand grip strength | Hand grip strength(right) |
| HT | 50 | 485,495 | Body size measures | Standing height |
| MTCIM | 20023 | 481,747 | Reaction time | Mean time to correctly identify matches |
| NTS | 20127 | 389,695 | Mental health | Neuroticism score |
| PEF | 3064 | 443,673 | Spirometry | Peak expiratory flow |
| PR | 4194 | 165,111 | Arterial stiffness | Pulse rate |
| SBP | 4080 | 455,787 | Blood pressure | Systolic blood pressure, automated reading |
| WC | 48 | 485,861 | Body size measures | Waist circumference |
| WT | 21002 | 485,276 | Body size measures | Weight |

**Supplementary Table 10.** Abbreviated names and details of the 4 traits in the pig dataset. The first column is the abbreviation, the second column is the number of records available for analysis, the third column indicates the category of the trait, and the last column contains the full names and specific details of these traits.

| Traits Abbr. | Count | Category | Description |
|---|---|---|---|
| BF | 75,538 | Carcass trait | Backfat thickness adjusted to 100 kg |
| AGE | 75,981 | Growth trait | Age adjusted to 100 kg |
| FCR | 19,822 | Feed efficiency trait | Feed conversion ratio |
| LMA | 51,785 | Carcass trait | Loin muscle area adjusted to 100 kg |

## Supplementary References

1. Bycroft, C. *et al.* The UK Biobank resource with deep phenotyping and genomic data. *Nature* **562**, 203-209 (2018).
2. Church, D.M. *et al.* Modernizing reference genome assemblies. *PLoS biology* **9**, e1001091 (2011).
3. Consortium, I.H. Integrating common and rare genetic variation in diverse human populations. *Nature* **467**, 52 (2010).
4. Chang, C.C. *et al.* Second-generation PLINK: rising to the challenge of larger and richer datasets. *Gigascience* **4**, s13742-015-0047-8 (2015).
5. Hof, J.P. & Speed, D. LDAK-KVIK performs fast and powerful mixed-model association analysis of quantitative and binary phenotypes. *Nature Genetics* **57**, 2116-2123 (2025).
6. Zhang, K. *et al.* AGIDB: a versatile database for genotype imputation and variant decoding across species. *Nucleic Acids Research* **52**, D835-D849 (2024).
7. Browning, B.L., Zhou, Y. & Browning, S.R. A one-penny imputed genome from next-generation reference panels. *The American journal of human genetics* **103**, 338-348 (2018).
8. Liu, Y. *et al.* ACAT: a fast and powerful p value combination method for rare-variant analysis in sequencing studies. *The American Journal of Human Genetics* **104**, 410-421 (2019).
9. Liu, Y. & Xie, J. Cauchy combination test: a powerful test with analytic p-value calculation under arbitrary dependency structures. *Journal of the American Statistical Association* **115**, 393-402 (2020).